%% file: paper_main.tex
\input{paper_preamble}

\begin{document}
\input{paper_body}
\end{document}

%% file: paper_preamble.tex
\documentclass[10pt,journal,compsoc]{IEEEtran}

\ifCLASSOPTIONcompsoc
  \usepackage[nocompress]{cite}
\else
  \usepackage{cite}
\fi
\usepackage{colortbl}
\usepackage{tikz}
\usepackage{xspace}
\usepackage{siunitx}
\usepackage{amsmath}
\usepackage{amssymb}
\usepackage[caption=false,font=footnotesize]{subfig}
\usepackage{soul}
\usepackage{textcomp}
\usepackage{graphicx}
\usepackage{wrapfig}
\usepackage{capt-of}
\usepackage{booktabs}
\usepackage{xcolor}
\usepackage{CJKutf8}
\usepackage[hidelinks]{hyperref}
\AtBeginDocument{%
  }

\input{macros}

\graphicspath{{imgs/}}

\title{\newproj: Advancing Real-Time Foveated Neural Rendering via Foveation-Aware Pruning and Stereo Warping}
\author{Weikai~Lin and Yu~Feng%
\IEEEcompsocitemizethanks{%
\IEEEcompsocthanksitem \parbox[t]{0.42\textwidth}{W. Lin is with University of Rochester, Rochester, NY, USA. E-mail: wlin33@ur.rochester.edu.\\
Y. Feng is with Shanghai Jiao Tong University, Shanghai, China. E-mail: y-feng@sjtu.edu.cn. Corresponding author: Y. Feng.}
}}

%% file: macros.tex
\newcommand*\circled[2]{\protect\tikz[baseline=(char.base)]{
            \protect\node[shape=circle,fill=black,inner sep=1pt] (char) {\textcolor{#1}{{\footnotesize #2}}};}}

\ifx\figurename\undefined \def\figurename{Figure}\fi
\renewcommand{\figurename}{Fig.}
\renewcommand{\paragraph}[1]{\textbf{#1} }

\newcommand{\Sect}[1]{Sec.~\ref{#1}}
\newcommand{\Fig}[1]{Fig.~\ref{#1}}
\newcommand{\Tbl}[1]{Tbl.~\ref{#1}}
\newcommand{\Eqn}[1]{Eqn.~\ref{#1}}

\newcommand{\proj}{\textsc{MetaSapiens}\xspace}
\newcommand{\newproj}{\textsc{MetaSapiens v2}\xspace}

\newcommand{\mode}[1]{\underline{\textsc{#1}}\xspace}

\newcommand{\RNum}[1]{\uppercase\expandafter{\romannumeral #1\relax}}

\newcommand{\cL}{\mathcal{L}}
\newcommand{\cM}{\mathcal{M}}

%% file: paper_body.tex
\bstctlcite{IEEEBST:etal}
\maketitle
\input{abs}

\input{intro}
\input{background}
\input{comp_aware_3dgs}
\input{hvs_aware_3dgs}

\input{opacity_enhanced_3dgs}

\input{binocular}
\input{hw}
\input{setup}
\input{eval}
\input{related}
\input{conclusion}

\begingroup
\let\oldthebibliography\thebibliography
\let\endoldthebibliography\endthebibliography
\renewenvironment{thebibliography}[1]{%
  \oldthebibliography{#1}%
  \scriptsize
  \setlength{\itemsep}{0pt}%
  \setlength{\parsep}{0pt}%
  \setlength{\parskip}{0pt}%
}{\endoldthebibliography}
\bibliographystyle{IEEEtran}
\bibliography{references, refs-cicero}
\endgroup

\begin{IEEEbiography}[{\includegraphics[width=1in,height=1.25in,clip,keepaspectratio]{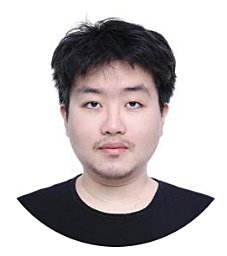}}]{Weikai Lin}
is a Ph.D. student in Computer Science at the University of Rochester, advised by Prof. Yuhao Zhu. His research spans algorithms and hardware, with a focus on AR/VR graphics, computer architecture, imaging systems, and artificial intelligence. He has published papers in major international venues including ASPLOS, ISCA, SIGGRAPH Asia, DAC, and CVPR. He received the ASPLOS 2025 Best Paper Award.
\end{IEEEbiography}

\begin{IEEEbiography}[{\includegraphics[width=1in,height=1.25in,clip,keepaspectratio]{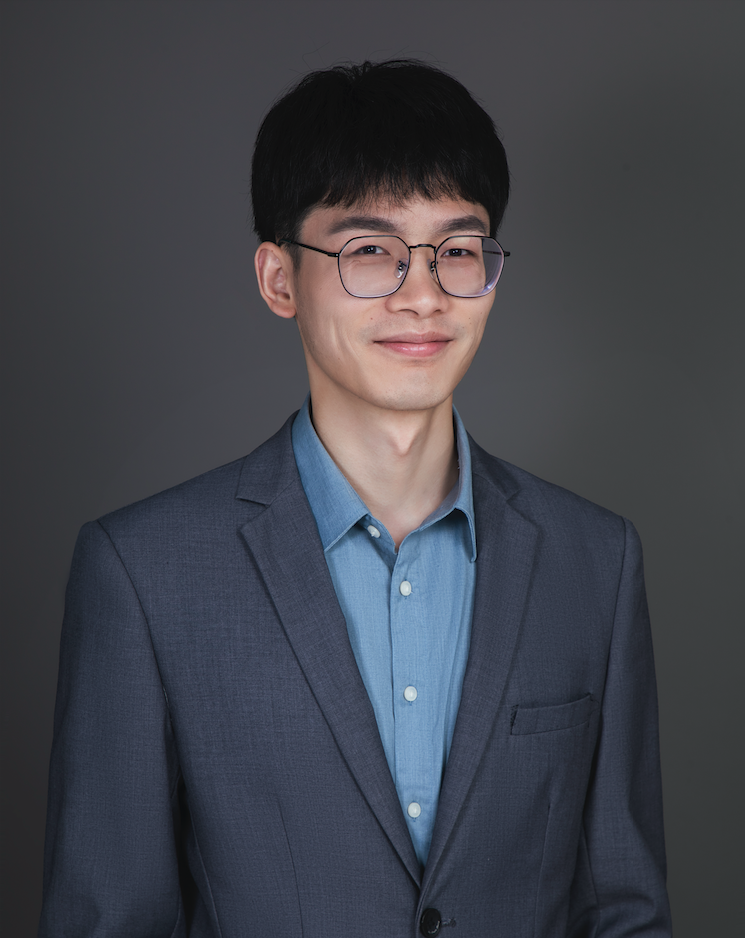}}]{Yu Feng}
is an Assistant Professor and Ph.D. advisor at the John Hopcroft Center, Shanghai Jiao Tong University. He received his Ph.D. in Computer Science from the University of Rochester in 2023 and was a postdoctoral researcher there from 2023 to 2024. He was selected for a national high-level young talent program in 2024 and the Young Development Incentive Program of the Computer Architecture Technical Committee in 2025. His research interests include computer architecture, edge inference acceleration, and edge-cloud collaborative computing. He has published more than 30 papers in ISCA, MICRO, ASPLOS, HPCA, and related venues, including 20 papers in the four major architecture conferences. His honors include the IEEE VR 2022 Best Paper Honorable Mention, ICCD 2024 Best Paper Nomination, ASPLOS 2025 Best Paper Award, and ASPLOS 2026 Best Paper Nomination.
\end{IEEEbiography}

%% file: abs.tex
\begin{abstract}
Point-Based Neural Rendering (PBNR) is emerging as
a promising class of rendering techniques, which are permeating all aspects of society, driven by a growing demand for real-time, photorealistic rendering in AR/VR and digital twins. 
However, achieving real-time PBNR on VR/AR devices is challenging.
This paper proposes {\newproj}, a PBNR system that delivers real-time neural rendering on  VR/AR devices while maintaining human visual quality.
{\newproj} combines four techniques.
First, we present an efficiency-aware pruning technique to optimize rendering speed. 
Second, we introduce a Foveated Rendering (FR) method with an efficient primitive for PBNR, leveraging humans' low visual acuity in peripheral regions to relax rendering quality and improve rendering speed.
Third, we leverage the redundancy between the two eyes and propose a selective warping method to further reduce the computation overhead in AR/VR binocular rendering.
Finally, we propose an accelerator design for binocular FR, addressing the load imbalance issue in (FR-based) PBNR and supporting warping for efficient binocular rendering.
Our evaluation shows that  \newproj achieves an order of magnitude speedup over existing PBNR models while maintaining the visual quality. 
\end{abstract}

%% file: intro.tex
\section{Introduction}
\label{sec:intro}

Rendering is becoming essential across domains.
For example, it powers digital twins in smart cities, healthcare, and telepresence.
Similarly, digital museums such as Virtual Smithsonian~\cite{vsmith} photorealistically render cultural heritage artifacts to provide immense humanity value for those who could not physically visit the museum.
Reinvigorated interest in Augmented and Virtual Reality (AR/VR) continues to elevate the demand for real-time, photorealistic rendering.

Point-Based Neural Rendering (PBNR), i.e.,the family of Gaussian Splatting algorithms~\cite{Kerbl2023GaussianSplatting, fan2023lightgaussian, lee2024compact, fang2024mini, lin2025metasapiens, huang2025seele, lin2025powergs}, is emerging as a compelling rendering paradigm.
It builds on classic point-based rendering techniques~\cite{zwicker2001surface}, revitalized through modern neural rendering methods~\cite{mildenhall2021nerf}.
Similar to Neural Radiance Fields (NeRF), PBNR achieves photorealistic rendering by learning scene radiance from images. 
However, it delivers significantly faster performance by replacing NeRF’s compute-heavy multilayer perceptrons (MLPs) with efficient point-based rasterization.

\begin{table*}[t]
\centering
\small
\caption{Contributions breakdown between \proj and \newproj.}
\label{tab:metasapiens_techniques}
\renewcommand{\arraystretch}{1.25}
\begin{tabular}{p{0.20\textwidth}p{0.74\textwidth}}
\hline
\textbf{Version} & \textbf{Contributions} \\
\hline
\proj & Perception-guided efficiency-aware pruning (\ref{sec:prune}); Foveated PBNR (\ref{sec:fr:mot}--\ref{sec:fr:train}); Hardware support for FR (\ref{sec:hw:load}) \\
\hline
\newproj & Efficient FR primitives (\ref{sec:fr:enhanced_prim}); Binocular rendering with selective warping (\ref{sec:bino}); Hardware support for selective binocular warping (\ref{sec:hw:warp}) \\
\hline
\end{tabular}
\vspace{-10pt}
\end{table*}

Nevertheless, PBNR still falls short of real-time performance on mobile devices, typically achieving under 10 frames per second (FPS) on the mobile Volta GPU~\cite{xaviersoc}.
This challenge becomes more severe in AR/VR scenarios, where binocular rendering is required for stereo vision, which effectively double the workload.
To address this, we introduce \newproj, a real-time foveated PBNR framework designed for efficient binocular rendering.

To further advance the performance frontier established by \proj~\cite{lin2025metasapiens}, \newproj introduces two key algorithmic contributions: a more efficient rendering primitive that eliminates the artifacts in foveated rendering and a binocular rendering pipeline that reduces redundant computations between the two eyes while maintaining perceptual quality.
Table~\ref{tab:metasapiens_techniques} summarizes the relationship between \proj and \newproj.
Specifically, \newproj consists of the following four key components.

\paragraph{Perception-Guided Efficiency-Aware Pruning.}
Recent PBNR accelerations have focused on pruning~\cite{fan2023lightgaussian, lee2024compact, fang2024mini}, which reduces model size but yields limited speedup.
This is because existing methods focus on minimizing the number of points, without considering their actual computational cost.
We observe that different points contribute unevenly to the total rendering workload.
Based on this insight, we propose an efficiency-aware pruning method that directly minimizes rendering/inference speed (\Sect{sec:prune}).
Furthermore, to ensure that the pruned model meets the required visual quality, our pruning method incorporates a human visual system (HVS) model to guide the process.
Specifically, we use the HVS model to monitor perceptual quality during training and iteratively prune the model as long as the quality remains above a perceptual threshold.

\paragraph{Foveated PBNR.}
\newproj also leverages the non-uniform characteristics of human vision to improve performance (\Sect{sec:fr:mot}--\Sect{sec:fr:train}).
Human visual acuity decreases in the periphery~\cite{wandell1995foundations}, a property long exploited in graphics through foveated rendering (FR)~\cite{guenter2012foveated}, where rendering quality is progressively reduced with increasing pixel eccentricity (i.e., as pixels are farther from the visual center) without compromising perceptual quality.

We introduce a FR method into PBNR (\Sect{sec:fr:mot}--\Sect{sec:fr:rep}) that gradually reduces the number of points used for rendering as pixel eccentricity increases.
Naively, one could train separate PBNR models for different eccentricity regions.
However, this approach incurs significant overhead that outweighs the speed benefits and also increases the model size.
To address this, we design a data representation in which points at higher eccentricities form strict subsets of those at lower eccentricities.
This hierarchical structure enables parameter and computation sharing across different eccentricity regions, improving rendering efficiency while reducing storage requirements.

While performance is critical, maintaining high visual quality is equally important.
To achieve this, we propose a training strategy  that uses quality-alignment to jointly guide both pruning and peripheral quality relaxation in FR (\Sect{sec:fr:train}).
Our method explicitly models human visual perception across different eccentricities and ensures quality consistency across the visual field.
As a result, the perceived quality matches that of the dense, non-foveated baseline.

\paragraph{Efficient FR Primitives.}
Through FR, our method significantly reduces the number of points required for rendering peripheral regions.
However, at high eccentricity, the points could become too sparse to fully cover the scene, leading to noticeable background leakage (black pixels) as shown in \Fig{fig:blackhole_overview}, which degrades visual quality and imposes a limit to the pruning ratio.
The root cause lies in PBNR’s original Gaussian primitives: although each primitive has a scalar opacity, its effective rendering alpha is obtained by multiplying this opacity with the Gaussian falloff (\Eqn{eqn:final_alpha}). As a result, alpha decays rapidly from the center outward.
This property makes it challenging to achieve full scene coverage with a limited number of Gaussian primitives.
To address this, we enhance the Gaussian primitives with a new alpha distribution function that controls the falloff and enables scene coverage with constrained gaussian primitives (\Sect{sec:fr:enhanced_prim}).
Our new distribution function enables better visual quality due to more efficient scene coverage.

\paragraph{Binocular Rendering with Selective Warping.}
In AR/VR, systems often render two separate frames for left and right eyes to provide stereo vision, which doubles the computational cost.
To leverage the view similarity between two closely coupled eyes, we propose a binocular rendering pipeline with selective warping to reduce the computation redundancy between these two eyes (\Sect{sec:bino}).
Specifically, we adopt a warping-based approach, which first renders an image for one eye and warps this image for the other eye.
To preserve visual quality, we propose selective warping, which applies warping only beyond a certain eccentricity, where human visual sensitivity is lower. We set this threshold by measuring the HVS quality (HVSQ) after warping and ensuring it matches the high-quality \proj, whose quality is shown by a user study to be comparable to a state-of-the-art dense PBNR (\Sect{sec:eval:sub}).

\paragraph{Architectural Support.}
While the techniques above already deliver an order-of-magnitude GPU speedup, we co-design an accelerator architecture to further boost performance and energy efficiency (\Sect{sec:hw}).
Beyond providing hardware support for FR, the architecture tackles a key bottleneck in PBNR, which is further exacerbated by FR: low hardware utilization caused by workload imbalance across tiles in a frame.
We address this with (1) a dynamic tile-merging scheme that balances workloads across tiles and (2) incremental pipelining of adjacent stages via line-buffering (\Sect{sec:hw:load}).
Moreover, to enable depth extraction and selective warping for the proposed binocular rendering, we extend the hardware with a lightweight depth-blending and warping unit (\Sect{sec:hw:warp}).

\paragraph{Result.}
We evaluate our method through subjective human studies  and objective performance and quality metrics.
In a user study (\Sect{sec:eval:sub}) with 12 participants, rendering quality of \proj is statistically indistinguishable from that of Mini-Splatting-D~\cite{fang2024mini}, a state-of-the-art PBNR.
For \newproj, we align its HVSQ with \proj by tuning the eccentricity threshold for warping (\Sect{eval:ecc_thresh}), ensuring that it maintains a similar perceptual-quality as \proj.
Against five state-of-the-art PBNR baselines, \newproj delivers higher objective quality (up to +0.49 dB PSNR) and faster rendering (up to $8.7\times$ on a mobile Volta GPU and $30.9\times$ with hardware support).
We summarize contributions of \proj and \newproj in \Tbl{tab:metasapiens_techniques}.

%% file: background.tex
\section{Background}
\label{sec:bck}
We first introduce the necessary background in PBNR (\Sect{sec:bck:pbnr}), followed by the main characteristics of the Human Visual System (HVS) and how they are exploited by Foveated Rendering to improve rendering speed (\Sect{sec:bck:hvs}).
We then describe stereo vision and binocular rendering in AR/VR (Sec.~\ref{sec:bck:binocular}), outlining their perceptual basis, the computational cost, and the redundancy between stereo views that can be reduced through warping.

\begin{figure}[t]
    \centering
    \includegraphics[width=0.83\columnwidth]{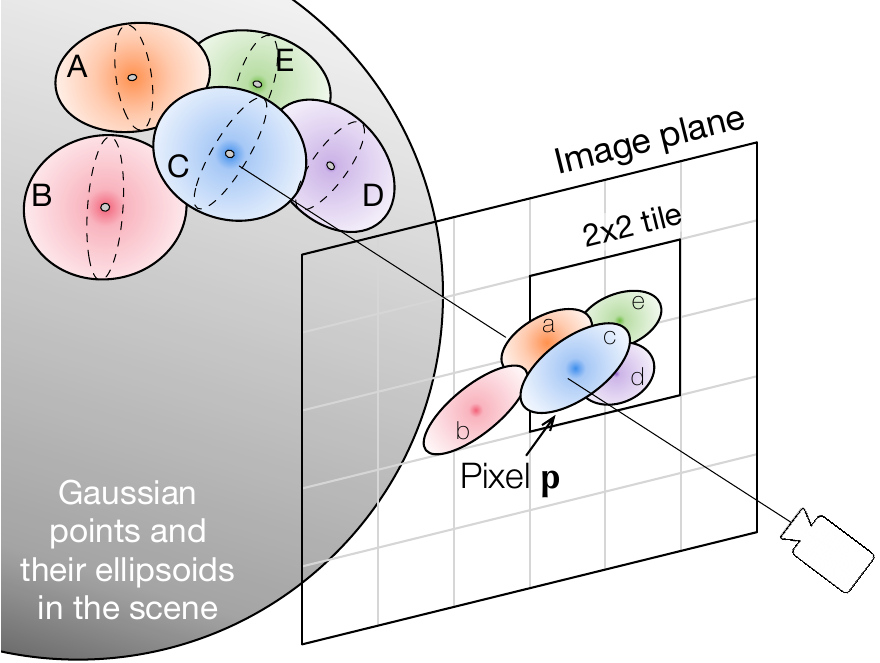}
\caption{
Illustration of PBNR, which parameterizes the scene with a set of points, each associated with a 3D Gaussian distribution that gives rise to an ellipsoid.
The ellipsoids are projected to ellipses on the image plane, where the ellipses are sorted (per tile, e.g., $2\times 2$ pixels).
The color of a pixel is calculated by integrating the contribution of each intersecting ellipse (e.g., \texttt{a}, \texttt{c}, \texttt{d}, \texttt{e} for \textbf{p}).
}
    \label{fig:pbnr}
    \vspace{-10pt}
\end{figure}

\subsection{Point-Based Neural Rendering}
\label{sec:bck:pbnr}

PBNR is a class of neural rendering techniques, exemplified by the 3D Gaussian Splatting (3DGS) algorithm~\cite{Kerbl2023GaussianSplatting} and its descendants~\cite{fan2023lightgaussian, lee2024compact, fang2024mini}.
Compared to previous neural rendering techniques, a.k.a., the NeRF-family algorithms~\cite{mildenhall2021nerf, muller2022instant}, PBNR is fundamentally more efficient (e.g., usually over 1,000 times faster), because it parameterizes the scene with discrete points (rather than voxels) to avoid redundant computations and renders via a lightweight rasterization-based process called splatting~\cite{zwicker2001surface} rather than the heavy MLP inference.
We use 3DGS as a running example for the PBNR pipeline (\Fig{fig:pbnr}). A trained model represents the scene as discrete Gaussian points, each associated with a 3D ellipsoid with trainable scale, position, orientation, opacity, and SH-based color. These parameters are learned offline from posed images via reconstruction loss. Online rendering then proceeds through \textit{Projection}, \textit{Sorting}, and \textit{Rasterization}.

\underline{Projection.}~
Each ellipsoid is first projected/splatted to an ellipse on the image plane\footnote{We use ``points'', ``ellipses'', and ``ellipsoids'' interchangeably: there is a one-to-one mapping between them.}.
In the example of \Fig{fig:pbnr}, the ellipsoids \texttt{A}--\texttt{E} in the scene are splatted to ellipses \texttt{a}--\texttt{e} on the image plane.
The goal is to identify, for each pixel tile (e.g., $2\times 2$), which ellipses intersect with the tile and thus contribute to the pixel colors in the tile.

\underline{Sorting.}~ 
For each tile, we sort all the intersecting ellipses based on their depths to the image plane;
that way, closer ellipses can be integrated first
when calculating pixel colors.
For instance in \Fig{fig:pbnr}, ellipse \texttt{c} would be the closest.

\underline{Rasterization.}~
Finally, we calculate the intersections of all the ellipses in a tile with each pixel.
The color of a pixel $\textbf{p}$ is then computed using the classic volume rendering method~\cite{max1995volume_rendering}, which integrates the contribution of all the intersecting ellipses from near to far:

\begin{subequations}
\label{eqn:alphablending}
\begin{align}
\textbf{p} = \sum_{i=0}^{N-1} T_i \alpha_i c_i + T_N c_{\text{bg}}, ~~~ T_i = \prod_{j=0}^{i-1} (1 - \alpha_j )
\label{eq:p}\\
\alpha_i = f(\text{opacity}_i, ..., \text{pose, pixel position}) \label{eq:f}\\
c_i = g(\text{SH}_0, ..., \text{SH}_n, \text{pose}) \label{eq:g}
\end{align}
\end{subequations}

\noindent where $N$ is the number of ellipses intersecting $\textbf{p}$ (i.e., \texttt{a}, \texttt{c}, \texttt{d}, \texttt{e} in \Fig{fig:pbnr}). {The function $f$ first projects the $i^{th}$ 3D ellipsoid to the image plane under the current pose, evaluates the projected 2D Gaussian at pixel position of $\textbf{p}$, and multiplies the result by the trainable opacity to obtain $\alpha_i$. The function $g$ evaluates the spherical-harmonic coefficients of the same point under the current viewing direction to produce the view-dependent color $c_i$.}
The final transmittance term $T_N = \prod_{j=0}^{N-1}(1 - \alpha_j)$ represents the fraction of light that passes through all ellipses without being absorbed, and is used to weight the background color $c_{\text{bg}}$.
We refer readers to Kerbl et al.~\cite{Kerbl2023GaussianSplatting} for the details.

\begin{figure}[t]
    \centering
    \includegraphics[width=\columnwidth]{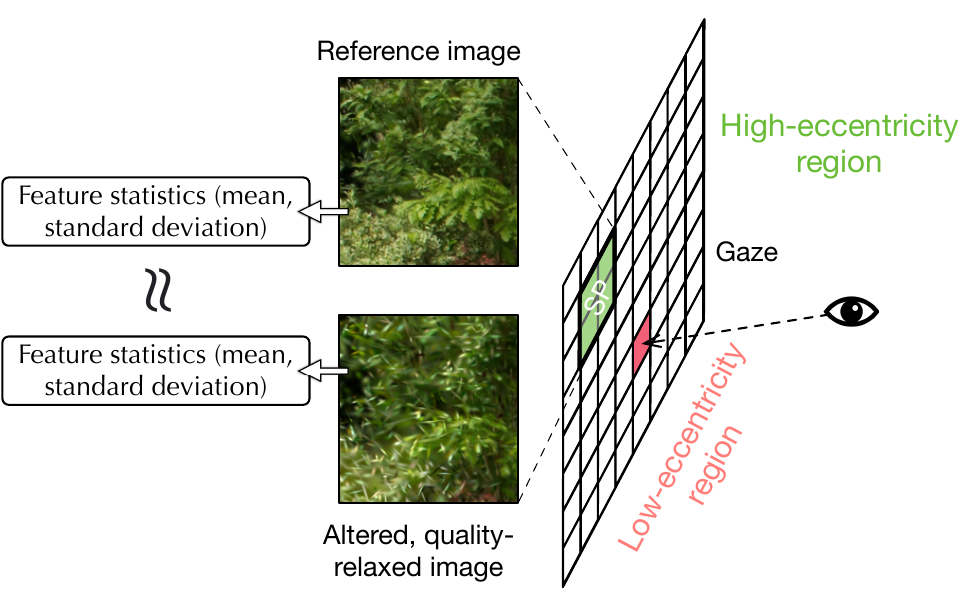}
\caption{
Pixels under the user's gaze have low eccentricities, where the human visual quality is the highest;
the peripheral pixels have high eccentricities where human visual acuity is low.
In peripheral regions, the visual stimulus (image) can be altered without being discriminable from the reference stimulus if the statistics of the image features are close, as quantified by the HVSQ metric (\Eqn{eq:hvsq}).
SP: spatial pooling.
}
    \label{fig:fr_example}
    \vspace{-10pt}
\end{figure}

\subsection{Human Visual System and Foveated Rendering}
\label{sec:bck:hvs}

\paragraph{Foveated Rendering.}
It is well-known that human visual acuity drops as eccentricity increases, i.e., when objects are placed more toward the visual periphery~\cite{walton2021beyond}.
This is due to a combination of larger pooling sizes~\cite{rodieck1985parasol} and a sparser photoreceptor distribution~\cite{Song:2011:ConeDensity} on the retina as the eccentricity increases.
Foveated Rendering (FR)~\cite{guenter2012foveated} exploits the fall-off in peripheral acuity to accelerate rendering by reducing quality in high-eccentricity regions, where visual changes are less noticeable, as illustrated in \Fig{fig:fr_example}.
While in classic FR the peripherial rendering quality is relaxed by lowering the resolution,
neural rendering offers another dimension: reducing the computational \textit{workload} of each pixel. 
This new dimension is possible as the rendering load of each pixel is controlled by inferencing a learned model, which offers many knobs for accuracy-vs-speed trade-offs that have been extensively studied~\cite{yang2019quantization}.
For instance, one can train a smaller model for rendering the visual periphery~\cite{deng2022fov}.
This paper will explore FR knobs unique to PBNR.

\paragraph{Modeling HVS.}
A key challenge in FR is determining how much quality can be relaxed without visible artifacts. Standard metrics such as PSNR and SSIM ignore the eccentricity-dependent acuity drop in HVS~\cite{walton2021beyond}, making them inadequate for FR: low peripheral PSNR may still be visually acceptable, as illustrated in \Fig{fig:fr_example}.

We instead use the eccentricity-aware HVS Quality (HVSQ) metric~\cite{walton2021beyond}. Given a reference image, an altered image, and per-pixel eccentricities determined by display resolution and eye-display distance, HVSQ measures their perceived similarity; lower is better. HVSQ is based on spatial poolings, where retinal responses aggregate over regions that grow with eccentricity, typically quadratically.
Two images are difficult to distinguish if their feature-space statistics, such as mean and standard deviation, remain close within each pooling, where the feature space consists of predefined human-sensitivity features.
Computationally, HVSQ is defined as:

\begin{equation}
  HVSQ = \frac{1}{N}\sum_{i=1}^{N}\Big[\big(\cM(\text{I}^{a}_i) - \cM(\text{I}^{r}_i)\big)^2 + \big(\sigma(\text{I}^{a}_i) - \sigma(\text{I}^{r}_i)\big)^2\Big]
  \label{eq:hvsq}
\end{equation}

\noindent where $N$ is the number of pixels in an image (each pixel has a unique spatial pooling), $\text{I}^{r}_i$ and $\text{I}^{a}_i$ denote the features of the $i^{th}$ spatial pooling in the reference and the altered image, respectively; $\cM$ denotes arithmetic mean, and $\sigma$ denotes standard deviation.
Intuitively, HVSQ averages the distance between feature statistics across all spatial poolings; larger peripheral poolings provide more flexibility to alter pixels while preserving reference statistics.

\subsection{Binocular Rendering in AR/VR}
\label{sec:bck:binocular}
Human depth perception relies on both monocular and binocular cues.
Monocular cues, such as  accommodation and motion parallax, can be perceived with a single eye.
Binocular cues require both eyes: by converging on the same object, each eye captures a slightly different image due to the interpupillary distance (IPD), creating \emph{binocular disparity} that the brain fuses into a single 3D perception.

\paragraph{Binocular Rendering.} 
To deliver immersive experiences, AR/VR systems reproduce the binocular depth cue by rendering two images per frame (one for each eye) from viewpoints separated by the user’s IPD.
While this enables realistic stereo vision, it roughly \textit{doubles} the per-frame rendering cost compared to monocular rendering, making it more challenging to sustain the high frame rates (e.g., 75-90,Hz~\cite{meta_quest_pro}) required for a smooth experience.

\paragraph{Warping-Based Inter-View Redundancy Reduction.}
To reduce the pressure of binocular rendering, we can exploit the fact that the two eye images often differ only by small disparities, particularly for distant or background content, which creates redundancy across the two rendering passes.
A classic approach to leverage this redundancy is \emph{depth-based image reprojection} (i.e. warping), which reuses rendered pixels from one view, reprojects them into the other using depth, and then fills (i.e. render) disoccluded pixels  (visible in the second view but not in the first)~\cite{fehn2004depth}.
However, this approach requires precise and dense scene depth for accurate warping.
Such depth is typically available in synthetic environments, however, it is difficult to obtain from real-world scenes used to train neural rendering models.

%% file: comp_aware_3dgs.tex
\section{Perception-Aware Pruning}
\label{sec:prune}

This section introduces a pruning framework for accelerating PBNR while meeting target perceptual quality.
We first identify the root cause of why existing pruning methods are not as effective as expected~(\Sect{sec:prune:perf}).
We then present two techniques to address this cause: intersection-aware pruning~(\Sect{sec:prune:metric}) and scale decay~(\Sect{sec:prune:scale}).
Finally, we describe how these techniques are integrated into an iterative pruning framework with perceptual quality guarantees~(\Sect{sec:prune:train}).

\begin{figure*}[t]
\centering
\begin{minipage}[t]{0.37\textwidth}
  \centering
  \includegraphics[width=\linewidth]{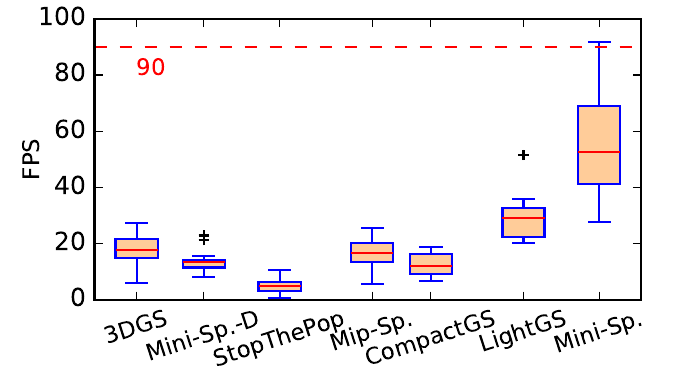}
  \caption{FPS distribution of recent PBNR models on common datasets measured on mobile Volta GPU on Jetson Xavier.}
  \label{fig:fps_boxplot}
\end{minipage}
\hspace{2pt}
\begin{minipage}[t]{0.3\textwidth}
  \centering
  \includegraphics[width=\linewidth]{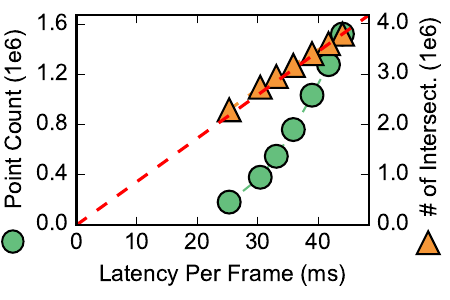}
\caption{Impact of point count and tile–ellipse intersections on per-frame latency.}
  \label{fig:model_vs_exec}
\end{minipage}
\hspace{2pt}
\begin{minipage}[t]{0.29\textwidth}
  \centering
  \includegraphics[width=\linewidth]{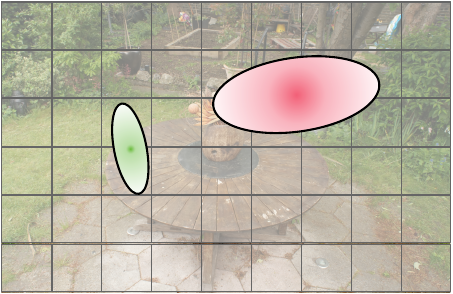}
  \caption{Two ellipses intersect different numbers of tiles, leading to different computation cost.}
  \label{fig:large_small}
\end{minipage}
\vspace{-10pt}
\end{figure*}

\subsection{Motivations}
\label{sec:prune:perf}

\paragraph{Speed.}
Recent PBNR models remain far from real time on mobile GPUs. \Fig{fig:fps_boxplot} reports FPS on three datasets~\cite{barron2022mip,Knapitsch2017,hedman2018deep}, measured on the Jetson Xavier Volta GPU across recent PBNR models~\cite{fan2023lightgaussian, lee2024compact, fang2024mini, Kerbl2023GaussianSplatting, yu2024mip, radl2024stopthepop}. Dense models are generally the slowest, while pruning-based models reduce model size but still fall short of AR/VR real-time targets of 75--90 FPS~\cite{meta_quest_pro}. Binocular rendering further exacerbates the challenge by requiring two views per pose.

\paragraph{Why is Existing Pruning Insufficient?}
Existing pruning methods focus on reducing the point count in a model, which is ineffective for improving speed in PBNR.
To quantify this, \Fig{fig:model_vs_exec} shows the inference latency ($x$-axis) vs. point count (left $y$-axis) of LightGS~\cite{fan2023lightgaussian} (which prunes 3DGS~\cite{Kerbl2023GaussianSplatting}) trained on the \texttt{bicycle} trace in the Mip-NeRF 360 dataset at different pruning levels (between 75\% and 97\%).
The latency reduction rate is slower than that of the point reduction rate.
The reason that reducing the point count is ineffective for acceleration is because the computational costs associated with different points vary.
\Fig{fig:large_small} shows the intuition, where there are two ellipses projected onto the image plane.
The smaller ellipse intersects with only two tiles whereas the larger one intersects with eight.
As a result, the larger one is used in calculating more pixel colors and is naturally responsible for more computation.

Therefore, what \textit{does} impact the inference speed is the number of tile-ellipse intersections.
\Fig{fig:model_vs_exec} shows the latency vs. the average number of intersections per tile (right $y$-axis) for each pruned LightGS model;
the latency reduction rate and intersection reduction rate match.

\subsection{Intersection-Aware Pruning}
\label{sec:prune:metric}

Our pruning aims to reduce tile--ellipse intersections with minimal quality loss. We define \textit{Computational Efficiency} (CE) to measure a point's pixel contribution per unit compute:
\begin{equation}
\text{CE}_i = \frac{\text{Val}_i}{\text{Comp}_i}.
\label{metric}
\end{equation}
Points with low CE consume substantial computation while contributing little, and are therefore pruned first.

$\text{Val}_i$ is the number of pixels dominated by point $i$, where a pixel is dominated by the point with the largest rasterization contribution $T_i\alpha_i$ in \Eqn{eq:p}. $\text{Comp}_i$ is the number of tiles intersecting and using point $i$, which directly affects rendering speed. Since CE depends on camera pose, we compute it for each training pose and use the maximum CE across poses, which is less biased than averaging. During pruning, we sort points by CE and remove the lowest-CE points, while controlling the quality as described in \Sect{sec:prune:train}.

\subsection{Scale Decay}
\label{sec:prune:scale}

Orthogonal to pruning, scale decay reduces tile--ellipse intersections by shrinking large ellipses that are frequently used during rendering. We define \textit{Weighted Scale} (WS) to weight each point's scale by its rendering usage:
\begin{equation}
\text{WS} = \frac{1}{N}\sum\limits_{i=0}^{N-1} \text{S}_i \text{G}_i,
\end{equation}
where $N$ is the number of points and $\text{S}_i$ is the scale of point $i$, defined as its maximum span. The usage weight $\text{G}_i$ is:
\begin{equation}
\text{G}_i = (\text{U}_i > T) \cdot ( \text{U}_i - T),
\end{equation}
where $\text{U}_i$ is the number of tiles using point $i$ and $T$ is a threshold. Thus, points used by fewer than $T$ tiles do not contribute, while  frequently used points are penalized more.
We integrate WS into training as an additional loss term:
\begin{equation}
\cL = \cL_\text{quality} + \gamma \cdot \text{WS},
\label{eq:sc-loss}
\end{equation}
where $\gamma$ controls the strength of scale decay.

\begin{figure}[t]
    \centering
    \includegraphics[width=0.48\textwidth]{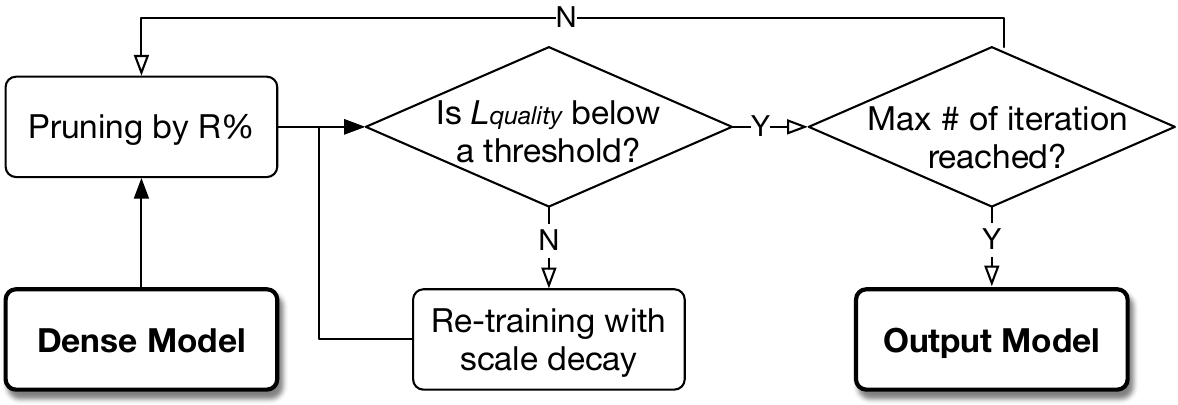}
    \caption{
    The procedure to obtain an efficient PBNR model given a dense model.
    We iteratively apply pruning and re-training with scale decay (guided by $\cL$ in \Eqn{eq:sc-loss}) while controlling for quality ($\cL_{quality}$).}
    \label{fig:mon_loss}
    \vspace{-10pt}
\end{figure}

\subsection{Putting It All Together: Perception-Guided Pruning}
\label{sec:prune:train}

Pruning and scale decay are conceptually orthogonal, but scaling an ellipse changes its CE, so they must be optimized jointly while satisfying the target perceptual quality. We therefore use an iterative procedure that combines CE pruning and scale decay, as shown in \Fig{fig:mon_loss}.

Starting from a dense model, we compute CE for all points and iteratively prune the lowest-CE fraction ($R=10\%$ in our implementation) until the quality loss $\cL_\text{quality}$ in \Eqn{eq:sc-loss} exceeds a threshold. We then fine-tune the pruned model with the composite loss $\cL$ to recover quality while applying scale decay, and resume pruning once the quality is met again. 
This repeats until the iteration budget is reached. $\cL_\text{quality}$ can be PSNR, SSIM, or another metric; We use an HVS-inspired metric for eccentricity-dependent acuity.

Our iterative procedure has two advantages.
First, it combines pruning and scale decay.
Second, it does not require quality-specific hyper-parameter tuning to achieve a specific 
visual quality: monitoring and controlling for $\cL_{quality}$ automatically yield a model at a given quality.

%% file: hvs_aware_3dgs.tex
\section{Foveated PBNR}
\label{sec:fr}

This section presents a Foveated Rendering (FR) method tailored for PBNR.
We begin with the core idea and associated challenges~(\Sect{sec:fr:mot}), followed by an efficient data representation that enables effective FR~(\Sect{sec:fr:rep}).
Next, we describe how to train FR models by leveraging the perception-guided iterative pruning introduced earlier~(\Sect{sec:fr:train}).
Finally, we discuss the background leakage issue at high eccentricity, which arises from sparse points, and introduce an enhancement for PBNR Gaussian primitives to allow efficient scene coverage~(\Sect{sec:fr:enhanced_prim}).

\subsection{Main Idea and Challenges}
\label{sec:fr:mot}

\begin{figure*}[t]
    \centering
    \includegraphics[width=0.88\textwidth]{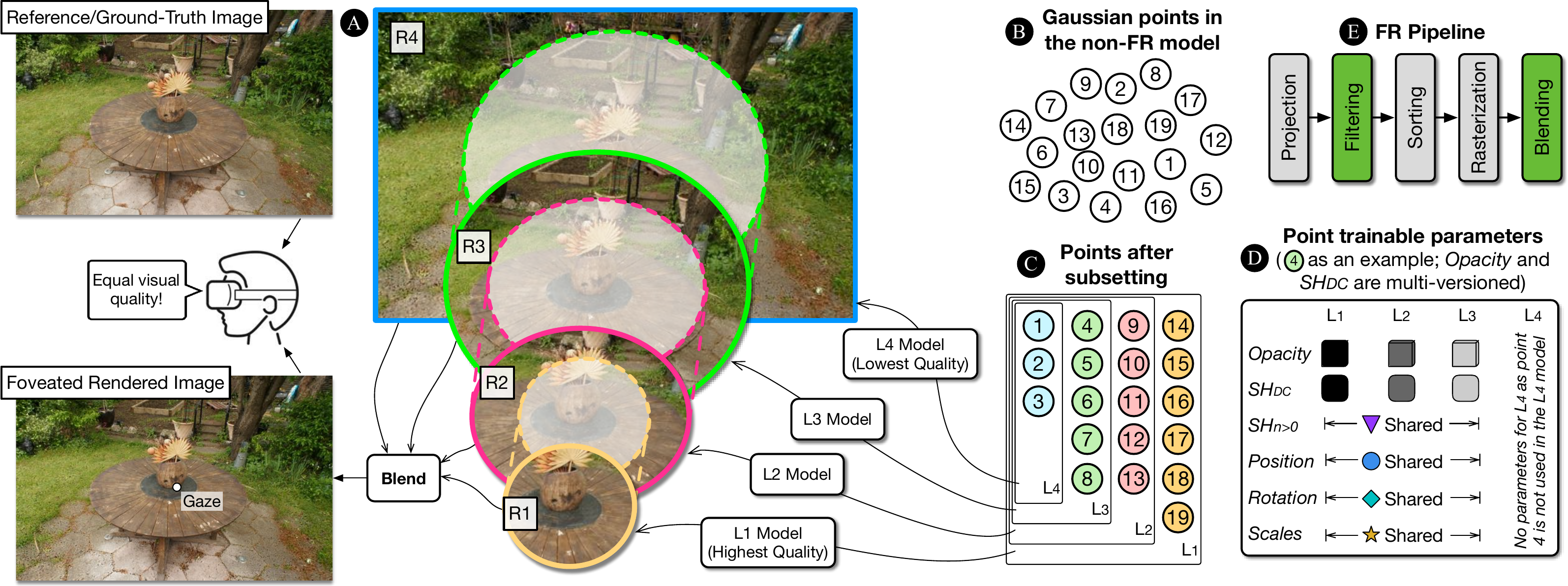}
    \caption{
    The general idea of FR for PBNR.
    \circled{white}{A}: We train multiple models (four in this example), each with a different quality and is responsible for rendering a different quality region in the image (\boxed{R1} -- \boxed{R4}).
    The four quality regions are blended together to generate the final image.
    The goal is for the FR-rendered image to have the same visual quality as the reference image (e.g., generated by a dense model) when judged by humans.
    \circled{white}{B}: Points in the original non-FR model.
    \circled{white}{C}: Our hierarchical point representation to support compute- and data-efficient FR.
    We subset the points so that points used to train a higher-level (lower quality) model are strictly a subset of that used by a lower-level model.
    The \textit{quality bound} $m$ of a point is the highest level that uses the point (e.g., $m=3$ for Point 4).
    \circled{white}{D}: To provide more flexibility for training, we selectively allow key trainable parameters to differ across levels; these parameters are the opacity of a point and the Direct Current (DC) component of the SH coefficients ($\text{SH}_\text{DC}$).
    Other (trainable) parameters of a point are shared across all the levels \textit{that use the point} (e.g., no parameter in $L_4$ for Point 4).
    \circled{white}{E}: The rendering pipeline augmented to support FR (augmentations in green).
    }
    \label{fig:fr_algo}
    \vspace{-10pt}
\end{figure*}

We accelerate rendering by relaxing the rendering quality at the visual periphery, leveraging the low peripheral visual acuity in HVS.
We illustrate the idea in \Fig{fig:fr_algo}, panel \circled{white}{A}.

\paragraph{Main Pipeline.}
Following prior FR work~\cite{deng2022fov, guenter2012foveated}, we divide an image into $N$ eccentricity regions, each rendered at a different quality level by a separate model. The gaze region uses the highest-quality model (\boxed{R1}), while peripheral regions use lighter models obtained through pruning and scale decay (\Sect{sec:prune}). A straightforward baseline is thus a multi-model design, where each region is rendered by its own model and boundary pixels are blended.

Panel \circled{white}{E} shows the FR-augmented pipeline with two new stages. First, after projection, we \textit{filter} each model's points outside its assigned quality region. Second, after rendering, we \textit{blend} overlapping boundary pixels to avoid sharp transitions between quality levels. As in prior FR methods~\cite{guenter2012foveated, deng2022fov}, each model renders slightly beyond its region boundary, so boundary pixels are rendered twice and interpolated for a smooth transition.
While this multi-model FR design is conceptually simple, it raises three challenges.

\paragraph{Challenge 1: Performance Overhead.}
FR can potentially accelerate rendering because it reduces the amount of rasterization work in low-quality regions.
However, it has two sources of performance overhead.

First, all $N$ models must go through the Projection and Filtering stages.
In our profiling, these two stages can take up to 18\% of the rendering time.
Second, blending also adds overhead.
Empirically we find that about 25\% of the pixels are to be blended and, thus, rendered twice.

\paragraph{Challenge 2: Storage Overhead.}
{Storing a separate model for every eccentricity band would multiply the already large memory footprint of PBNR. As a concrete example, the \texttt{bicycle} scene from Mip-NeRF 360~\cite{barron2021mip} occupies roughly 1.4 GB after standard 3DGS training~\cite{Kerbl2023GaussianSplatting}. Even after applying recent pruning techniques~\cite{fan2023lightgaussian}, the same scene remains about 490 MB, which is still a substantial burden for mobile AR/VR devices.}

We address the first two challenges using an efficient data representation, as we will discuss in \Sect{sec:fr:rep}.

\paragraph{Challenge 3: Controlling Quality.}
FR must be done in a way that guarantees human visual quality --- how do we decide the amount of relaxation at each level?
We describe a training strategy to guarantee consistent human visual quality across all levels, as described in \Sect{sec:fr:train}.

\subsection{Efficient FR Representation with Selective Multi-Versioning}
\label{sec:fr:rep}

{To reduce both runtime and storage costs, we design the FR models so that quality levels reuse the same underlying point set whenever possible. Specifically, each lower-quality model is formed from a strict subset of the points in the next higher-quality model. Panel \circled{white}{C} of \Fig{fig:fr_algo} shows this hierarchy after applying it to the original points in Panel \circled{white}{B}: $L_1$ retains the largest point set and therefore provides the best quality, whereas $L_4$ keeps the smallest set and provides the most relaxed quality.}

{This nested organization avoids replicating points across quality levels. The total point storage across all $N$ levels is therefore bounded by the largest model, $P_{total} = \text{max}_{i=1}^N{P_i} = P_1$, instead of the much larger independent-model total $\sum_{i=1}^N{P_i}$. Consequently, FR introduces no additional point-storage overhead, and its front-end compute overhead is also limited because Projection and Filtering run once on the shared hierarchy rather than independently for every quality model.}

{A point in this hierarchy may be shared by several models, from $L_1$ through $L_m$. We call $m$ the point's \textit{quality bound}, because levels above $m$ omit that point; for example, Point 4 in \Fig{fig:fr_algo} has $m=3$. At runtime, Projection assigns the point to a tile whose eccentricity maps to quality level $t$. The Filtering stage in Panel \circled{white}{E} then discards the point whenever $t > m$, so later stages process only points allowed at that tile's quality level.}

\paragraph{Selective Multi-Versioning.}
{Strict subsetting alone gives too little freedom to tune quality separately across eccentricity regions. If every trainable parameter is shared, a point contributes the same $\alpha_i c_i$ term in \Eqn{eq:p} regardless of which quality level renders it.}
{That fixed contribution is undesirable because the same point may fall into different quality regions as camera pose and gaze position change, and the color contribution should adapt to the region.}

{We therefore use selective multi-versioning, shown in Panel \circled{white}{D}. A point with quality bound $m$ keeps level-specific copies only for a small subset of trainable parameters, with one copy for each level that includes the point. In practice, we version four parameters: opacity and the three Direct Current SH coefficients ($\text{SH}_{DC}$), which have the strongest effect on pixel color in experiments. \Sect{sec:eval:fr} shows that this selective versioning is necessary to preserve visual quality.}

\subsection{HVS-Guided FR Training}
\label{sec:fr:train}

So far we have focused on performance; FR must also determine how much each peripheral model can be weakened while preserving subjective quality. We use the HVSQ metric from \Sect{sec:bck:hvs}, which compares a reference image and an altered image while accounting for eccentricity-dependent acuity. Although \Eqn{eq:hvsq} defines HVSQ over the full image, we apply it to each quality region by iterating only over spatial poolings in that region. During offline training, the reference is available from held-out images or dense-teacher renderings, allowing HVSQ to guide model derivation before deployment.
Our goal is to keep HVSQ consistent across quality levels and matched to the baseline. We first obtain the highest-quality $L_1$ model from a dense model using pruning and scale decay. We then derive $L_{i+1}$ from $L_i$ using the iterative pruning/retraining procedure in \Sect{sec:prune:train}, with two changes. First, we use region-specific HVSQ as $\cL_\text{quality}$ in \Eqn{eq:sc-loss} and match each level's HVSQ to that of $L_1$, ensuring consistent visual quality across the field of view. Second, we disable scale decay during this stage because ellipse scale is not a multi-versioned parameter.

\begin{figure*}[t]
\centering
\makebox[\textwidth][c]{%
\begin{minipage}[t]{0.33\textwidth}
  \vspace{0pt}
  \centering
  \includegraphics[width=\linewidth]{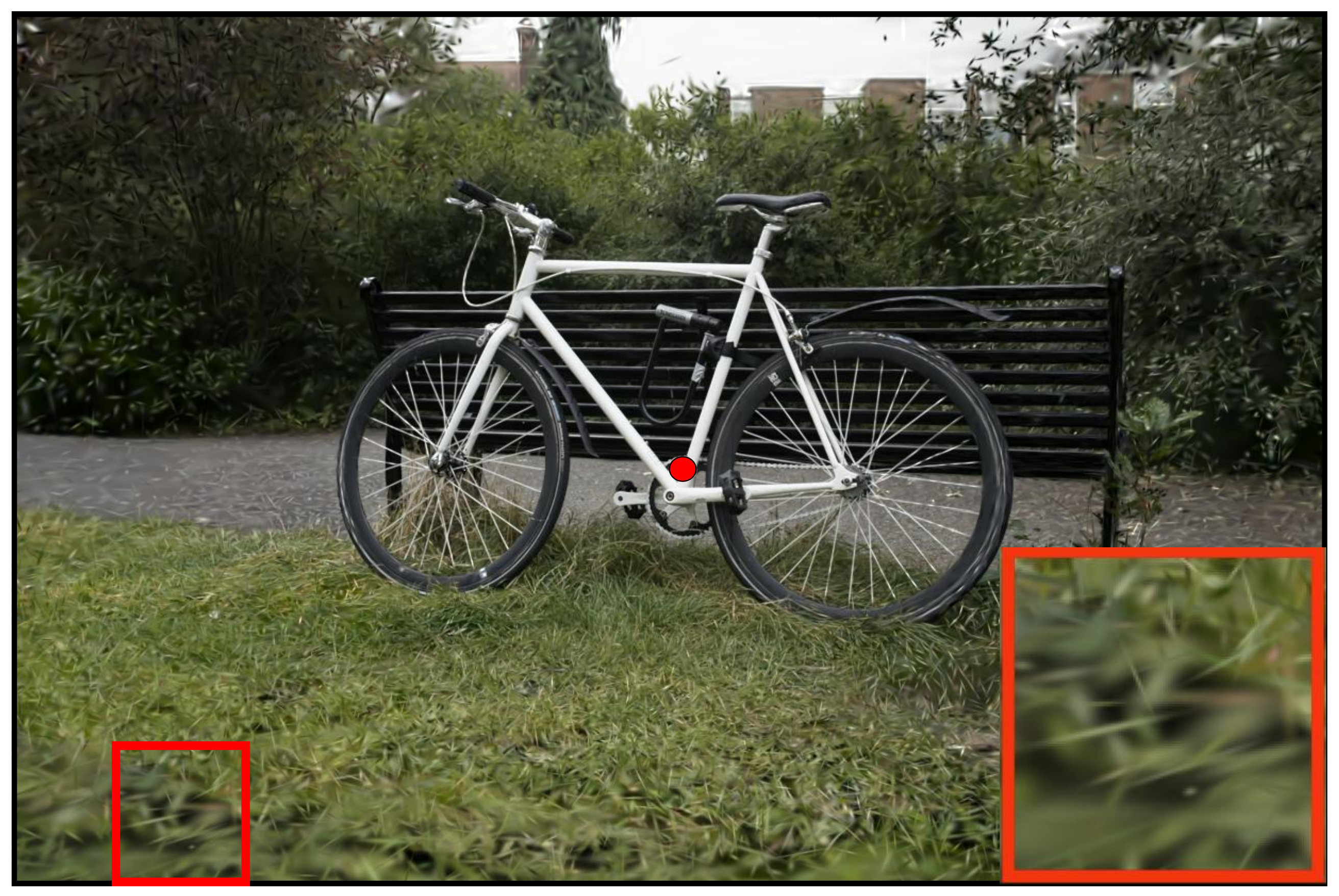}
  \captionof{figure}{
    Illustration of black hole artifacts produced by \proj-H in peripheral regions.
    The gaze is fixed at the center (red dot). We use \texttt{bicycle} from the Mip-NeRF 360 dataset.
  }
  \label{fig:blackhole_overview}
\end{minipage}\hfill%
\begin{minipage}[t]{0.62\textwidth}
  \vspace{0pt}
  \centering
  \begin{minipage}[t]{0.31\linewidth}
    \vspace{0pt}
    \centering
    \subfloat[$o_i = 1$, $\text{amp}_i = 1$\label{fig:falloff1}]
  {
      \includegraphics[width=\linewidth]{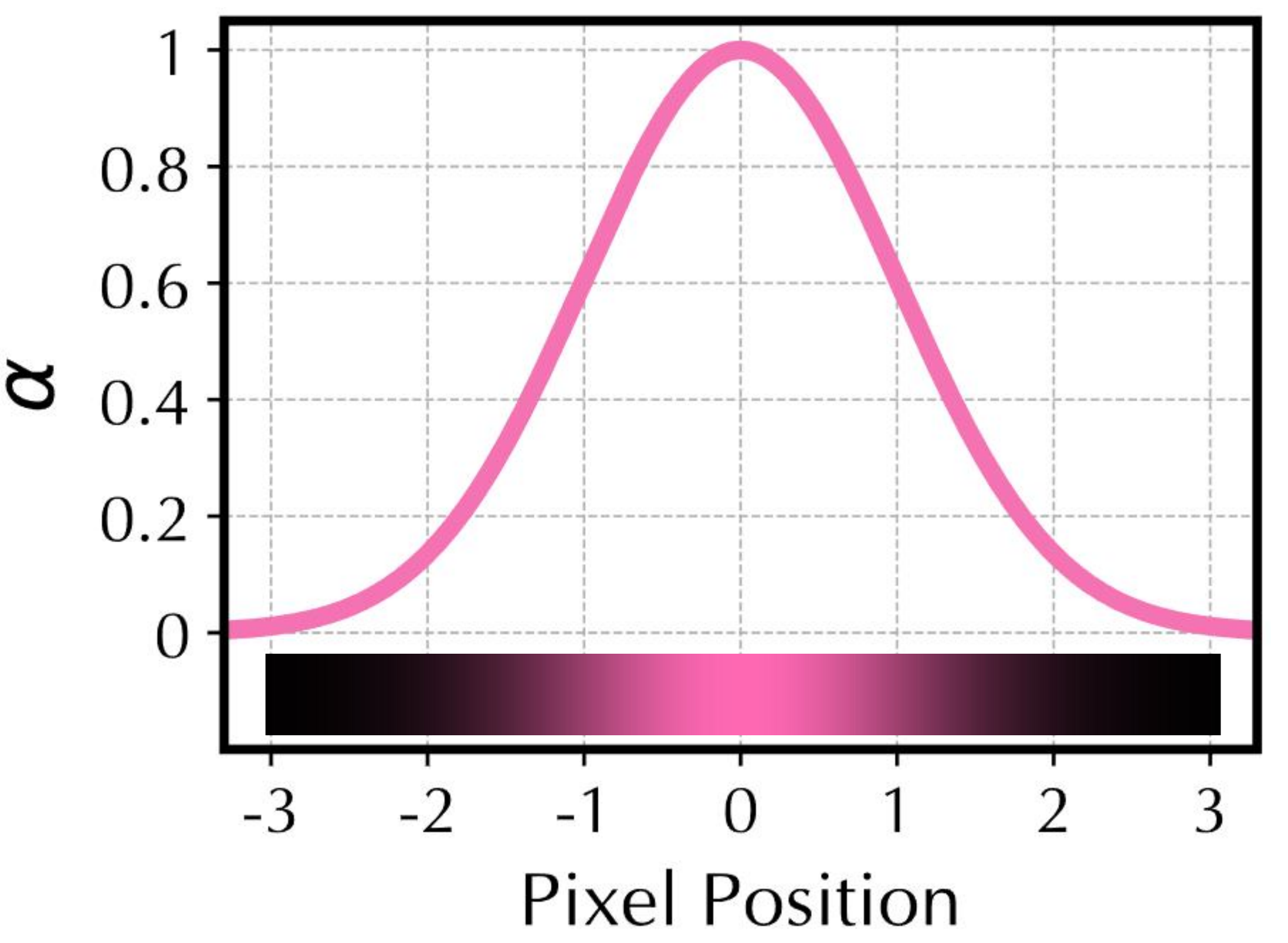}
  }
  \end{minipage}\hfill%
  \begin{minipage}[t]{0.31\linewidth}
    \vspace{0pt}
    \centering
    \subfloat[$o_i = 1$, $\text{amp}_i = 8$\label{fig:falloff8}]{
      \includegraphics[width=\linewidth]{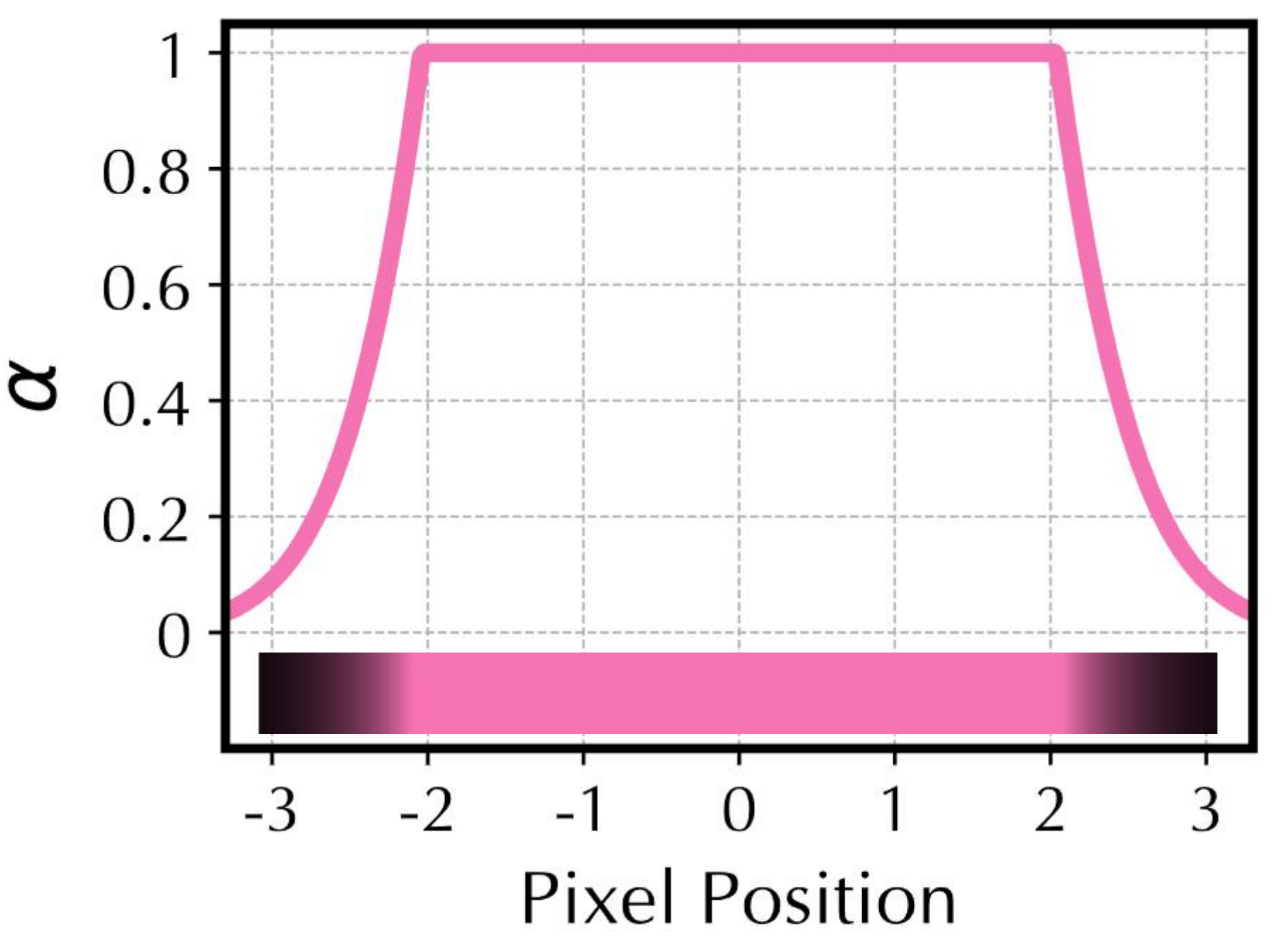}
  }
  \end{minipage}\hfill%
  \begin{minipage}[t]{0.31\linewidth}
    \vspace{0pt}
    \centering
    \subfloat[$o_i = 1$, $\text{amp}_i = 128$\label{fig:falloff128}]{
      \includegraphics[width=\linewidth]{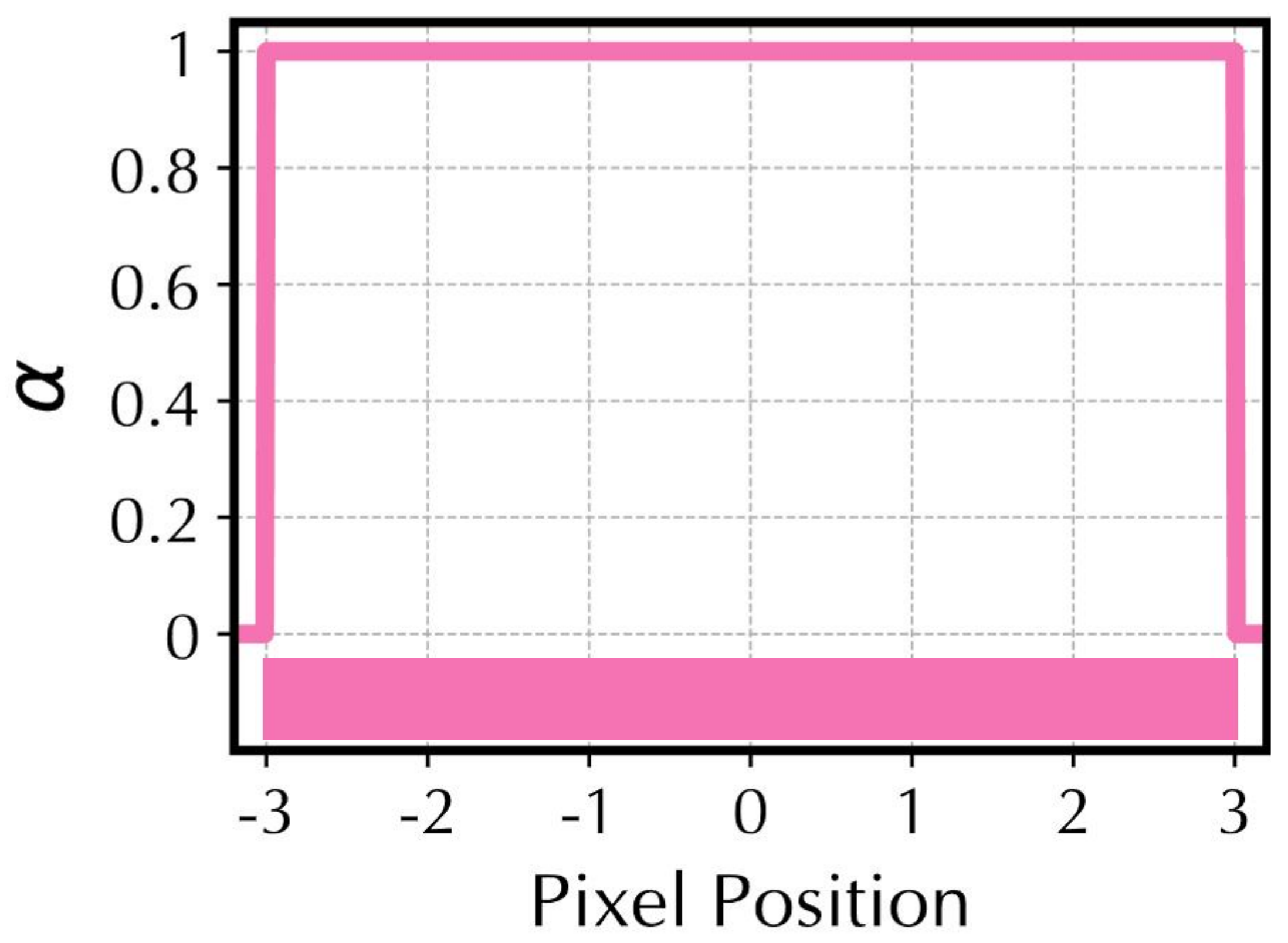}
  }
  \end{minipage}

\captionof{figure}{
Illustration of how the opacity amplifier $\text{amp}_i$ (\Eqn{eqn:amp}) expands Gaussian coverage (\Eqn{eqn:final_alpha}).  
We visualize the alpha values (top) and the blending results between a pink Gaussian and a black background (bottom).  
(a) With no amplification (original 3DGS), background leakage occurs at the edges due to limited alpha support. 
(b) With $\text{amp}_i = 8$, the Gaussian covers more pixels, reducing leakage.  
(c) With $\text{amp}_i = 128$, the Gaussian fully covers all intersected pixels, effectively eliminating background leakage.
}

  \label{fig:ampeffect}
\end{minipage}
}
\vspace{-10pt}
\end{figure*}

%% file: opacity_enhanced_3dgs.tex
\subsection{Opacity-Enhanced Primitives for Efficient Scene Representation}
\label{sec:fr:enhanced_prim}

Our FR method reduces peripheral point count and accelerates PBNR, but aggressive pruning causes a sharp HVSQ drop due to \emph{background leakage}. As shown in \Fig{fig:blackhole_overview}, sparse peripheral Gaussians may fail to cover the scene, producing visible black holes. This occurs because only a small set of Gaussians contributes to alpha blending in the periphery (\Eqn{eqn:alphablending}), and the original Gaussian opacity decays quickly from the center (\Fig{fig:falloff1}). Thus, pixels not near a Gaussian center may accumulate insufficient alpha, leading to holes.

\paragraph{Opacity Distribution Function.}
To mitigate background leakage, we introduce a learnable opacity amplifier $\text{amp}_i$ for each Gaussian $i$, allowing its effective opacity to exceed 1 and compensate for Gaussian falloff:
\begin{equation}
\label{eqn:amp}
o^\text{amp}_i = \text{amp}_i \cdot o_i,
\end{equation}
where $o_i \in (0,1)$ is the original opacity and $\text{amp}_i$ can be larger than 1. This increases the coverage of sparse peripheral Gaussians that would otherwise leave pixels blank.

Since alpha blending requires $\alpha_i \in [0,1]$, we clip the amplified alpha to avoid negative transmittance and color contributions:
\begin{equation}
\label{eqn:final_alpha}
\alpha^\text{amp}_{ij} = \min\left(1,\ o^\text{amp}_i \cdot \exp\left(-\frac{1}{2} (\mathbf{p}_j - \boldsymbol{\mu}_i)^\top \mathbf{\Sigma}_i^{-1} (\mathbf{p}_j - \boldsymbol{\mu}_i) \right)\right),
\end{equation}
where the exponential term is the Gaussian falloff at pixel $\mathbf{p}_j$ with center $\boldsymbol{\mu}_i$ and covariance $\mathbf{\Sigma}_i$~\cite{kerbl20233d}. The amplified opacity is thus a learnable pre-clipped factor that reshapes the alpha support before final blending. As shown in \Fig{fig:ampeffect}, large $\text{amp}_i$ values can flatten the falloff and cover more intersected pixels, reducing background leakage and enabling more aggressive peripheral pruning.
During training, we initialize $\text{amp}_i=1$, preserving the original 3DGS rendering, and update it only for high-opacity Gaussians with $o_i>0.8$. Opacity enhancement is applied only to peripheral models used in regions $>1$ (\Fig{fig:fr_algo}), where Gaussians are sparse and background leakage is more likely.

%% file: binocular.tex
\begin{figure*}[t]
\centering
\includegraphics[width=\textwidth]{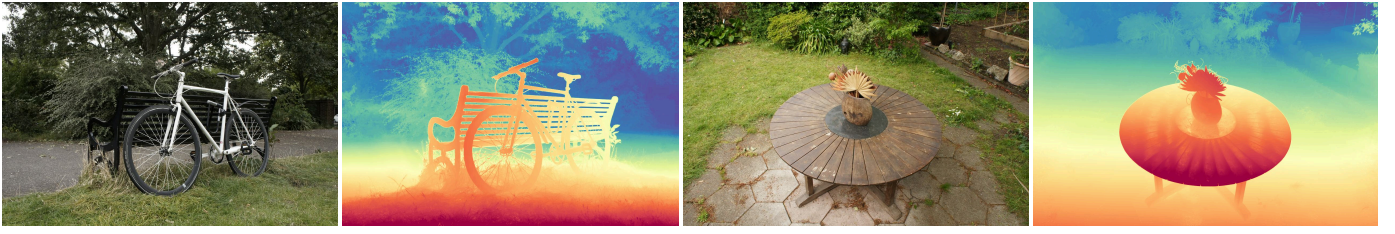}
\vspace{-10pt}
\caption{
Ground-truth RGB images and depth maps provided by depth blending. 
We use scenes with large spatial extent (\texttt{bicycle} and \texttt{garden}). 
It shows PBNR can provide reasonable depth map estimates for both foreground and background.
}
\label{fig:depth_map}
\vspace{-10pt}
\end{figure*}

\section{Selective Binocular Warping}
\label{sec:bino}

In this section, we propose a binocular rendering method for PBNR that leverages \emph{selective warping} to achieve speedup while preserving high quality.
We first present the motivation and challenges of applying warping in PBNR~(\Sect{sec:warp:mot}).
We then describe how to extract depth from PBNR rendering with minimal overhead~(\Sect{sec:warp:depth}). 
Finally, we propose a selective warping strategy that maintains quality while improving speed, even with noisy depth~(\Sect{sec:warp:method}).

\subsection{Motivation and Challenges}
\label{sec:warp:mot}
AR/VR systems must render two slightly different views to provide stereo cues (\Sect{sec:bck:binocular}), doubling the workload of PBNR, which is already far from real-time for monocular rendering~(\Sect{sec:prune:perf}).
Warp-and-fill can reduce this cost by reusing one eye's rendering for the other, but it requires dense and accurate depth.
Prior works~\cite{feng2024cicero, feng2024potamoi} rely on SfM meshes for depth, which are often incomplete, error-prone, and require extra storage and computation.

We address this by extracting full-frame depth directly during PBNR rendering (\Sect{sec:warp:depth}).
To tolerate depth errors, we selectively warp only pixels beyond an eccentricity threshold $\text{Ecc}_{\text{thresh}}$, where human visual sensitivity is lower (\Sect{sec:warp:method}).
This preserves the visual quality while reducing binocular rendering cost.

\subsection{Extracting Depth from PBNR Rendering}
\label{sec:warp:depth}

Inspired by depth-regularized PBNR training~\cite{li2024dngaussian}, where Gaussian depths correlate with scene geometry, we estimate per-pixel depth by alpha-blending the depths of contributing Gaussians, analogous to RGB blending (\Eqn{eqn:alphablending}). \Fig{fig:depth_map} shows example depth maps.
Depth extraction adds negligible overhead because per-Gaussian depth is already computed during sorting (\Sect{sec:bck:pbnr}), and alpha-blending weights are already available from RGB blending. Thus, each Gaussian only requires one extra multiply--accumulate (MAC) to contribute to pixel depth. On Nvidia Jetson AGX Xavier~\cite{xaviersoc}, depth extraction increases rendering time by only 5\%. This produces a dense full-frame depth map for warping, though noisy PBNR-estimated depth may cause geometric distortions, which \Sect{sec:warp:method} addresses.

\subsection{Eccentricity-Based Selective Warping}
\label{sec:warp:method}
As discussed in \Sect{sec:warp:depth}, PBNR-estimated depth can be noisy. We therefore use selective warping for high-quality binocular rendering: only one eye is rendered, and its RGB results are warped to the target eye using the estimated depth. To handle disocclusions, we track a valid mask during reprojection, fill small holes with median filtering, and rerender only tiles that still contain invalid pixels from the target-eye pose.
To tolerate noisy depth, we exploit the HVS's lower sensitivity to peripheral spatial distortions (\Sect{sec:bck:hvs}) and apply warping only beyond an eccentricity threshold $\text{Ecc}_{\text{thresh}}$, leaving the fovea unwarped. We choose the lowest threshold that meets a target quality $\text{HVSQ}_{\text{target}}$, set to match our high-quality FR baseline \proj~\cite{lin2025metasapiens}, thereby maximizing the warping ratio (\Sect{eval:ecc_thresh}).

%% file: hw.tex
\section{Hardware Support}
\label{sec:hw}
\begin{figure*}[t]
\centering
\includegraphics[width=\textwidth]{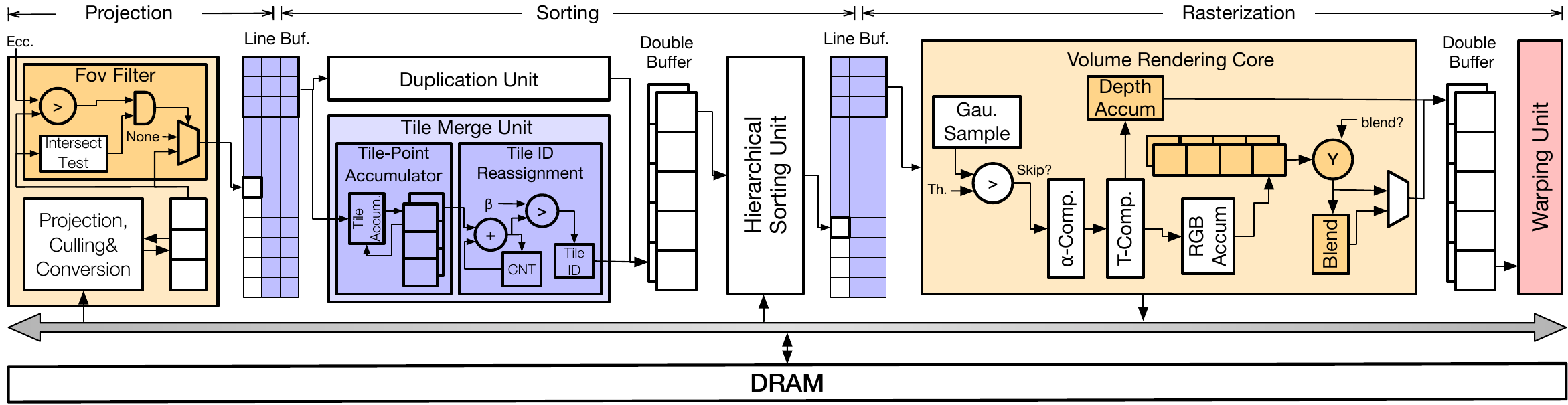}
\caption{
The overall architecture design. 
The the basic pipeline is similar to that of GSCore~\cite{lee2024gscore}, a recent PBNR accelerator.
We augment the baseline to support FR (yellow-colored) and warping (pink-colored), and to address the workload imbalance issue in PBNR/FR (blue-colored).
}
\label{fig:arch}
\vspace{-5pt}
\end{figure*}

\begin{figure*}[t]
  \centering
\begin{minipage}[c]{.48\textwidth}
  \centering
  \subfloat[Heatmap showing the number of intersections per tile in \texttt{bicycle}.]{%
      \label{fig:imbalance_exp}%
      \includegraphics[width=.47\linewidth]{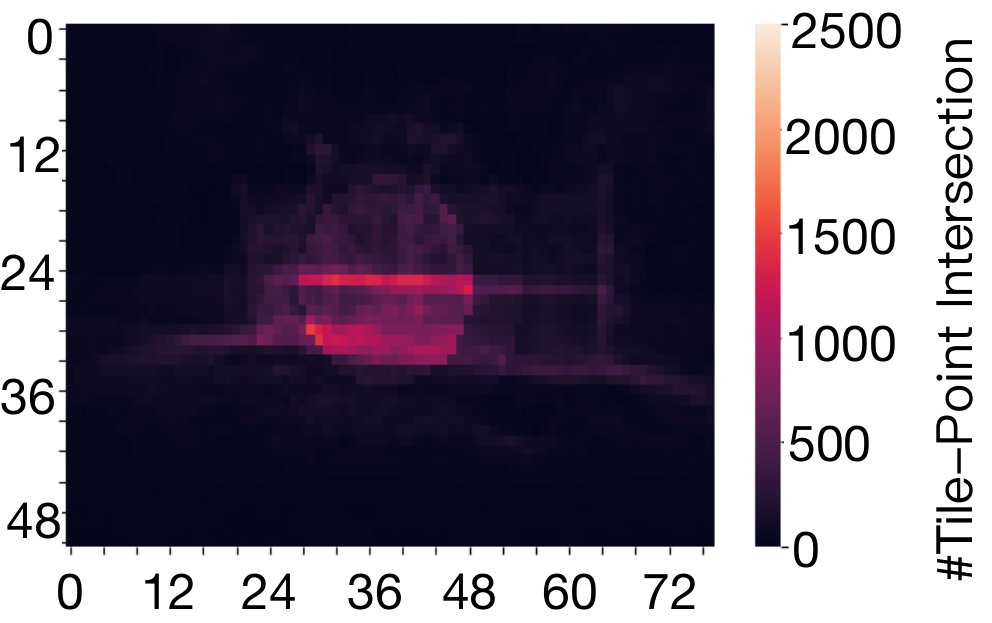}}%
  \hfill
  \subfloat[Boxplot of the intersection distribution in 5 traces (clipped at 1,500).]{%
      \label{fig:imbalance_across_scene}%
      \includegraphics[width=.47\linewidth]{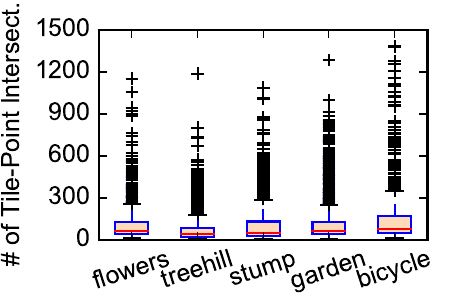}}
  \captionof{figure}{Workload imbalance, quantified by the number of intersections per tile, on the Mip-NeRF 360 dataset~\cite{barron2022mip}. 
  The top and bottom notches in the boxplot represent data points that are 1.5 Interquartile Range (IQR) above the third quartile and below the first quartile, respectively.}
  \label{fig:acc}
\end{minipage}
  \hfill
  \begin{minipage}[c]{.48\textwidth}
    \centering
    \includegraphics[width=1\linewidth]{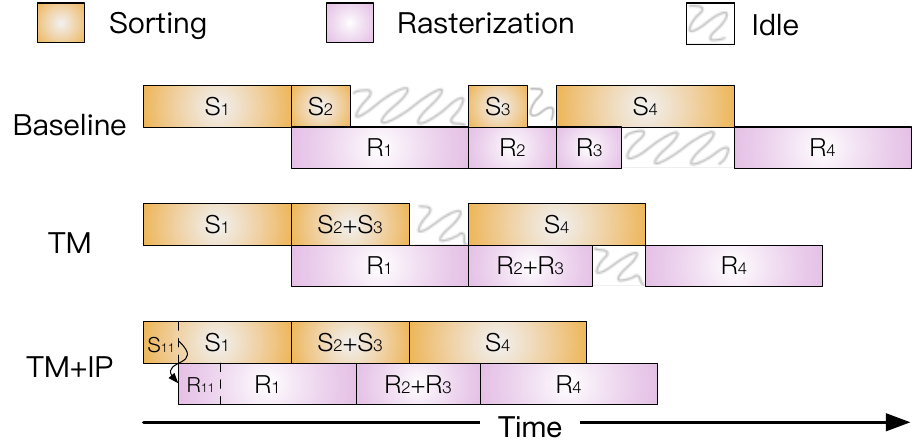}
    \captionof{figure}{The baseline faces frequent stalls when the workload is imbalanced across tiles.
    Tile Merging (TM) and Incremental Pipelining (IP) mitigate the workload-imbalanced issue and improve the pipelining efficiency.
    With IP, when the first sub-tile $S_{11}$ in the $S_1$ tile is available by the Sorting stage it can be processed by the Rasterization stage.}
    \label{fig:merged_pipeline}
  \end{minipage}
  \vspace{-10pt}
\end{figure*}
Complementing pruning and FR techniques, we propose an accelerator to further improve the performance.
We first provide an overview (\Sect{sec:hw:ov}) and discuss how the hardware addresses the low hardware utilization issue in PBNR, which is exacerbated by FR (\Sect{sec:hw:load}).
Finally, we introduce our hardware augmentation for depth extraction and warping in PBNR (\Sect{sec:hw:warp}), which enables the binocular rendering scheme proposed in \Sect{sec:warp:method}.

\subsection{Overview}
\label{sec:hw:ov}

Our architecture is built on top of GSCore~\cite{lee2024gscore}, a recent PBNR accelerator without FR.
The basic architecture is designed to support the three PBNR stages discussed in \Sect{sec:bck:pbnr}.
The three stages are pipelined across different tiles in a frame.
\Fig{fig:arch} shows the pipelined architecture, with the colored components denoting our augmentations.
The top panel in \Fig{fig:merged_pipeline} illustrates the pipelining process (omitting the Projection stage for simplicity).

\paragraph{Supporting FR.}
{We add hardware support for the two FR stages highlighted in green in Panel \circled{white}{E}; the corresponding blocks are shown in yellow in \Fig{fig:arch}. The first addition is a lightweight filter in Projection. For each projected point-tile entry, the filter checks the point's quality bound $m$ against the tile's current quality level $t$ and suppresses entries whose bound does not permit rendering at that level. The second addition is a Rasterization-side blending unit for the boundary regions described in \Sect{sec:fr:mot}. It keeps a small temporary pixel buffer and interpolates the two colors produced by neighboring quality-level models.}

\subsection{Addressing Load Imbalance}
\label{sec:hw:load}

\paragraph{The Issue.}
{Although the FR additions provide the required functionality, they expose a utilization problem: work is highly uneven across tiles. The pipeline schedules work at tile granularity (\Fig{fig:merged_pipeline}), and the cost of a tile is mainly determined by how many tile--ellipse intersections it sends to Sorting and Rasterization. \Fig{fig:imbalance_exp} visualizes these intersections for 16$\times$16 tiles on a Mip-NeRF 360 frame rendered by our four-level FR model. The count spans more than three orders of magnitude, with dense work near the gaze point and much lighter work in the periphery because peripheral regions use pruned models. \Fig{fig:imbalance_across_scene} confirms that the same pattern appears across Mip-NeRF 360 traces~\cite{barron2022mip}.}

{This uneven work distribution causes pipeline bubbles. The top panel of \Fig{fig:merged_pipeline} illustrates the effect by showing Sorting and Rasterization progressing through four tiles with different workloads. We mitigate this problem with two mechanisms, tile merging and incremental pipelining, whose hardware blocks are colored blue in \Fig{fig:arch}.}

\paragraph{Tile Merging.} 
{Tile merging balances the pipeline by grouping low-work tiles before Sorting. The Tile Merge Unit (TMU) performs this grouping inside the Sorting stage: it accumulates intersection counts from incoming tiles and combines adjacent tiles while their cumulative count remains below the threshold $\beta$.}
{The second panel of \Fig{fig:merged_pipeline} gives an example in which the second and third tiles are grouped, producing a more balanced work item and reducing stalls.}

{The TMU implements this policy with a two-stage counter pipeline. In the first stage, each Gaussian point updates the counter for its tile ID, and the intermediate counts are held in a temporary buffer. Once a tile's count is finalized, the second stage streams tiles in order, accumulates their intersection counts, and compares the running total with the merge threshold $\beta$. When the threshold boundary is reached, the TMU emits one merged work item: every original tile keeps its native tile ID and is also tagged with the merged-tile ID consumed by the sorting unit.}

\paragraph{Incremental Pipelining.}
{Tile merging narrows the workload gap across tiles, but it cannot make every merged tile equally expensive; residual differences in intersection count still leave some pipeline stalls.}

{To improve pipeline utilization further, we pass partial tile results between adjacent stages instead of waiting for a whole tile to finish. The baseline accelerator uses double buffers between stages, which forces the consumer to idle until the producer completes an entire tile. We instead partition a tile into smaller sub-tiles and allow the consumer to start as soon as a sub-tile becomes available. This is valid because pixel work is independent within a tile, and it is analogous to superpipelining in processor design~\cite{shen2013modern}. The final panel of \Fig{fig:merged_pipeline} illustrates how this incremental handoff reduces stalls.}

{We realize this handoff by replacing the inter-stage double buffers with line buffers (LBs)~\cite{hegarty2014darkroom}. Each LB is composed of small SRAM rows that store only the produced pixel rows needed by the next stage. With a 16$\times$16 tile, the consumer can begin once the required 16 rows for a sub-tile are present in the LB. Because the buffer stores sub-tile data not complete tiles, its capacity remains small.}
\subsection{Supporting Warping}
\label{sec:hw:warp}
Supporting warping in \newproj is straightforward: we add a warping unit after the volume-rendering core. Since rendering proceeds tile-by-tile (16$\times$16) in row-major order, we use a double buffer to overlap rasterizing one eye with warping the other. Each buffer stores 16 pixel rows, and the warping unit processes pixels row-by-row because warped pixels stay on the same row:
\begin{align} \begin{cases} y' & = y, \nonumber \\ x' & = x + \frac{Bf}{D}, \nonumber \\ \end{cases} \end{align}
where $\langle x,y\rangle$ and $\langle x',y'\rangle$ are the original and warped coordinates, $B$ is the eye distance, $f$ is the focal length, and $D$ is depth. A row depth buffer resolves conflicts when multiple pixels warp to the same target pixel. Disoccluded tiles are rerendered using the existing hardware.

%% file: setup.tex
\section{Experimental Setup}
\label{sec:exp}

\paragraph{FR Training Procedure.}
We use four quality regions following \proj~\cite{lin2025metasapiens}, starting at 0\textdegree, 18\textdegree, 27\textdegree, and 33\textdegree, and covering about 13\%, 17\%, 21\%, and 49\% of pixels, respectively. We use \mode{Mini-Splatting-D}~\cite{fang2024mini}, the current best-quality model, as the dense PBNR baseline. The $L_1$ model is obtained from the dense model using pruning and scale decay (\Sect{sec:prune:train}) for 50,000 iterations, followed by 5,000 iterations of HVSQ fine-tuning. Each lower-quality model is then derived from its immediate higher-quality model (\Sect{sec:fr:train}) with 10,000 iterations. Training takes roughly 3$\times$ longer than the dense model, mainly because HVSQ loss uses an open-source Python implementation, which could be accelerated with a CUDA implementation.

\paragraph{Variants.}
We offer a wide range of quality–speed trade-offs by designing three variants: \mode{\newproj-H}, \mode{\newproj-M}, and \mode{\newproj-L}, with progressively lower rendering quality and higher speed.
The $L_1$ model in these variants is pruned to achieve 99\%, 98\%, and 97\% of the PSNR of the dense model, respectively.
For each variant, we provide both monocular (\mode{Mono.}) and binocular (\mode{Bino.}) versions.
For our binocular variant, we set the eccentricity threshold for selective warping to 18\textdegree~to balance quality and speedup, which is justified in \Sect{eval:ecc_thresh}.

\paragraph{Datasets.}
We evaluate three real-world datasets: Mip-NeRF360~\cite{barron2022mip}, Tanks \& Temples~\cite{Knapitsch2017}, and DeepBlending~\cite{hedman2018deep}, totaling 13 traces.
Camera poses for these datasets are estimated by using COLMAP~\cite{schonberger2016structure} as SfM method.

Since the scales from COLMAP are not absolute and cannot be directly linked to real-world user interpupillary distance (IPD).
To perform binocular rendering, we first align COLMAP scales to real-world metric scale using MoGe-2~\cite{wang2025moge}, a geometry foundation model that provides metric depth. 
We then set the translation between the two eyes to 6.3 cm for binocular rendering, which corresponds to the average human IPD.

\paragraph{Hardware Implementation.}
We develop a RTL implementation of the accelerator, where the basic pipeline (the uncolored in \Fig{fig:arch}) is similar to GScore~\cite{lee2024gscore},
with the resource allocation adjusted for a more balanced pipeline for our workloads (8 Culling and Conversion Units, a single Hierarchical Sorting Unit, and a 16$\times$16 Volume Rendering Core array).
Our RTL design is implemented via Synposys synthesis and Cadence layout tools in TSMC 16nm FinFET technology.
Each line buffer has a capacity of 1 KB, and the double buffer before the sorting unit is 64 KB.
The SRAMs are generated by an Arm compiler.
The DRAM is modeled after four channels of Micron 16 Gb LPDDR3-1600 memory~\cite{micronlpddr3}
Our hardware evaluation is based on an RTL design rather than a taped-out chip.

Overall, we have an area of 2.74 mm$^2$.
The volume Rendering Core takes 63\% of the total area, other stages occupy the rest 30 \%; the SRAMs comprise 7\% of the total area.
The area of the warping unit is negligible. 
Our area is larger than that of GScore (1.45 mm$^2$), whose area is scaled to 16nm using the DeepScaleTool~\cite{sarangi2021deepscaletool}. 
We will show in \Sect{sec:eval:gscore} that we outperform GScore even under the same area when the latter is scaled up.

\paragraph{Baselines.}
We group evaluated baselines into four categories. Dense PBNR baselines render every pixel with one full-quality model. 
Pruned PBNR baselines reduce model size at the cost of quality but do not use gaze-contingent rendering. 
Prior \proj variants use perception-guided pruning and the FR hierarchy without opacity enhancement and selective binocular warping in \newproj.
We compare against ten recent PBNR models:
\begin{itemize}
    \item Dense PBNR models: \mode{3DGS}~\cite{Kerbl2023GaussianSplatting}, which is the earliest PBNR,  and \mode{Mini-Splatting-D}~\cite{fang2024mini}, \mode{Mip-Splatting}~\cite{yu2024mip}, \mode{StopThePop}~\cite{radl2024stopthepop}, which are state-of-the-art work that improve upon \mode{3DGS}.
    \item Pruned PBNR models: \mode{LightGS}~\cite{fan2023lightgaussian}, \mode{CompactGS}~\cite{lee2024compact}, and \mode{Mini-Splatting}~\cite{fang2024mini}. The first two are pruned from \mode{3DGS} and the last one is from \mode{Mini-Splatting-D}.
    \item FR PBNR models: \mode{\proj-H}, \mode{\proj-M}, and \mode{\proj-L}~\cite{lin2025metasapiens}, which are our method without the opacity enhancement (\Sect{sec:fr:enhanced_prim}) and selective warping (\Sect{sec:bino}).
\end{itemize}

We compare our selective multi-versioning FR (\Sect{sec:fr:rep}) with two PBNR-based FR methods using the same quality regions. \mode{SMFR} (Single-Model FR) uses the dense $L_1$ model from \mode{\proj-H} and randomly samples points in lower-quality regions, corresponding to a strict-subset version of our representation without selective multi-versioning. \mode{MMFR} (Multi-Model FR)~\cite{deng2022fov} uses the same $L_1$ model, with each lower-quality model pruned independently from it. Both methods use the same number of points per level as \mode{\proj-H}.

{
\paragraph{User Study Procedure.}
{We do not conduct a new user study for the v2 opacity-enhancement or binocular-warping extensions. Instead, we rely on the prior \mode{\proj-H} user study to validate the subjective quality of the FR principle based on HVSQ alignment.
We then ensure the quality of \newproj by aligning its HVSQ quality to \mode{\proj-H}} (\Sect{eval:ecc_thresh}).
We assess the subjective quality of our HVSQ-based FR principle (\Sect{sec:fr:train}) through an IRB-approved user study with 12 participants, comparable in scale to prior work~\cite{deng2022fov}. Using four diverse scenes (\texttt{bicycle}, \texttt{room}, \texttt{drjohnson}, and \texttt{truck}), we compare \mode{\proj-H}, which applies only pruning-based HVSQ alignment, against the highest-quality baseline \mode{Mini-Splatting-D}. The goal is to verify that our FR principle introduces no noticeable perceptual degradation.

}

%% file: eval.tex
\section{Evaluation}
\label{sec:eval}
\begin{figure*}[t]
\centering
\begin{minipage}[t]{0.46\textwidth}
  \centering
  \includegraphics[width=\linewidth]{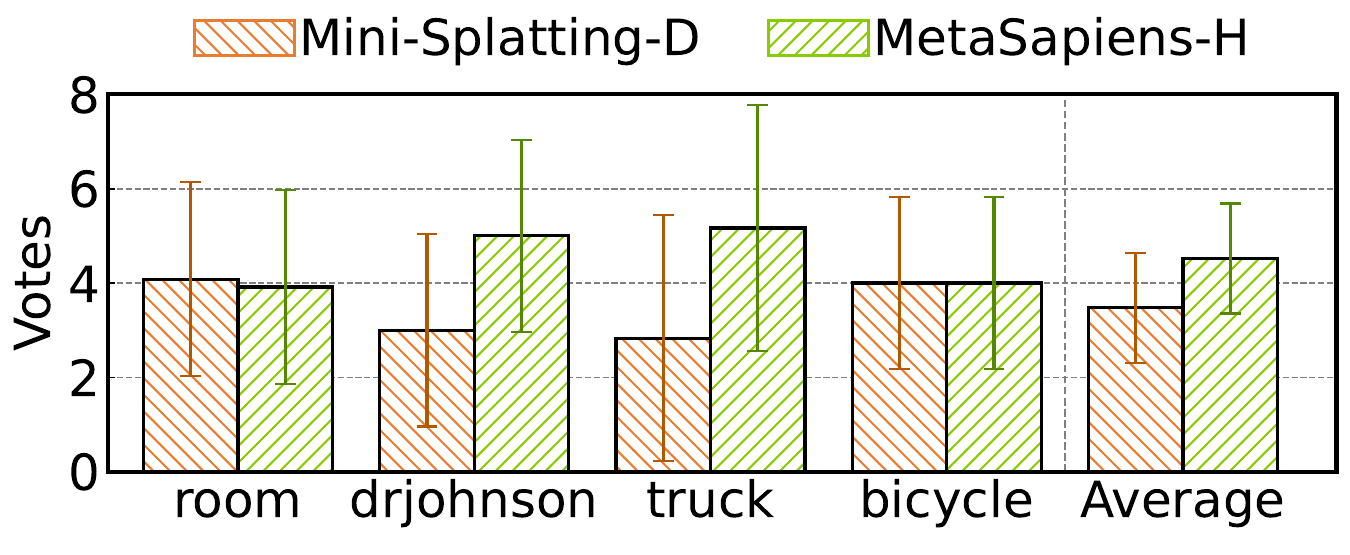}
  \caption{The average number of times the two methods are preferred by users (a tie would be 4-vs-4).
    Error bars indicate the standard deviation within the participants.
    Users either have no preference or prefer our method (binomial test on the average result; $p$ < 0.01).
}
  \label{fig:sub_exp}
\end{minipage}
\hspace{4pt}
\begin{minipage}[t]{0.37\textwidth}
  \centering
  \includegraphics[width=\linewidth]{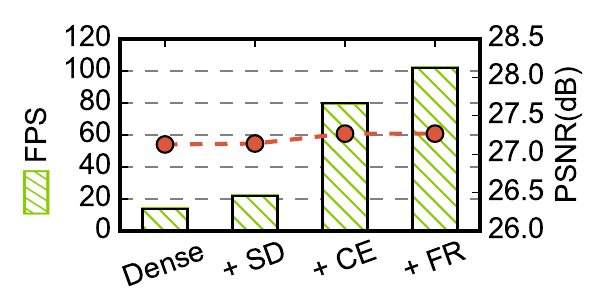}
  \caption{Ablation study teasing apart the impact of various techniques.
  The FPS results are obtained on Jetson Xavier and averaged over all traces.}
  \label{fig:fr_comp}
\end{minipage}
\vspace{-10pt}
\end{figure*}

\begin{figure*}[t]
\centering
\begin{minipage}[t]{0.98\textwidth}
  \centering
  \includegraphics[width=\linewidth]{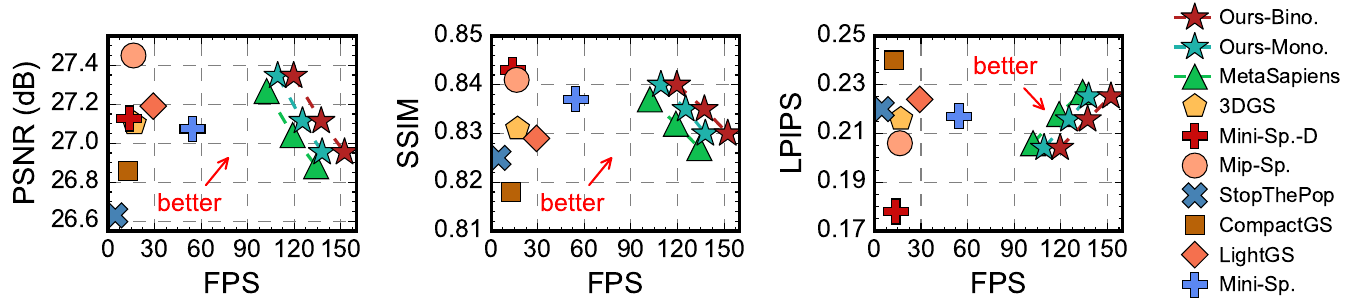}
  \caption{Performance and objective rendering quality (PSNR, SSIM, LPIPS) comparison across the seven baselines and the three \proj variants on the mobile Volta GPU.
   \mode{3DGS}, \mode{Mini-Splatting-D}, \mode{Mip-Splatting}, and \mode{StopThePop} are dense models, and the other three baselines are pruned models.
  }
  \label{fig:gpu}
\end{minipage}
\vspace{-10pt}
\end{figure*}

We first show that the subjective quality of our HVSQ-based FR principle (\Sect{sec:fr:train}) is  no worse than \mode{Mini-Splatting-D}, a state-of-the-art dense PBNR (\Sect{sec:eval:sub}). We then evaluate \newproj's mobile-GPU speed--quality trade-off (\Sect{sec:eval:render}), quantify the contributions of its main algorithmic modules (\Sect{sec:eval:ablation}), justify selective warping via HVSQ-based eccentricity-threshold selection (\Sect{eval:ecc_thresh}), and demonstrate additional hardware speedups and energy savings (\Sect{sec:eval:hw}). Finally, we compare against other FR methods (\Sect{sec:eval:fr}) and a prior PBNR accelerator (\Sect{sec:eval:gscore}).

\subsection{Subjective Evaluation of the FR Principle.} 
\label{sec:eval:sub}
{We perform a subjective experiment confirming that the HVSQ-aligned FR model \mode{\proj-H} (\Sect{sec:fr:train}) has no worse perceptual quality than \mode{Mini-Splatting-D}, a state-of-the-art dense PBNR baseline.}

\Fig{fig:sub_exp} shows the average number of participants who prefer the two methods for each video.
A tie would be 4-vs-4, since each video is watched eight times by each user.
We find that users either have no preference or prefer our method over \mode{Mini-Splatting-D}.
The results are statistically significant through a binomial test with $p$ < 0.01; the null hypothesis is ``users prefer \mode{Mini-Splatting-D} more than 50\% of the time''.

It might initially look surprising that we have equal or better subjective quality than \mode{Mini-Splatting-D}, a dense model from which we prune and build our FR model.
Further inspection and interviewing participants show two reasons.
First, our HVS-aware fine-tuning (\Sect{sec:exp}) better aligns the statistics of human-sensitive features with the ground truth.
Second, some points in the dense model are trained with inconsistent information across camera poses, leading to incorrect luminance changes over time; pruning those points helps alleviate this inconsistency.

{This result supports two uses of HVSQ in our evaluation. First, aligning the HVSQ of higher-eccentricity regions to the foveal region preserves subjective quality, validating HVSQ as the perceptual constraint for FR. Second, because \mode{\proj-H} is subjectively comparable to dense PBNR, later \newproj experiments can use \mode{\proj-H}'s HVSQ as the quality target for opacity enhancement and selective warping, rather than requiring a separate user study for each extension.}

\subsection{GPU Results}
\label{sec:eval:render}
We evaluate GPU performance on Nvidia Jetson AGX Xavier~\cite{xaviersoc}, a representative mobile platform for VR-like use cases. \Fig{fig:gpu} reports the average FPS over five runs for each camera pose and scene across all datasets. Monocular FPS measures single-eye images per second, while binocular FPS measures stereo frames per second with one directly rendered eye and one warped eye, including depth extraction, warping, and selective tile re-rendering/filling for disocclusions.
We compare \newproj variants with baselines using PSNR, SSIM, and LPIPS, which measure foveal-region quality and are commonly used in prior work. These metrics are computed using the foveal model ($L_1$), since HVSQ is already aligned across eccentricities. A prior \proj user study~\cite{lin2025metasapiens} shows that our FR principle (\Sect{sec:fr:rep}) achieves quality no worse than dense PBNR.

\begin{figure*}[t]
  \centering
  \subfloat[Point Count and FPS Improvement in R3]
  {
    \includegraphics[width=0.33\textwidth]{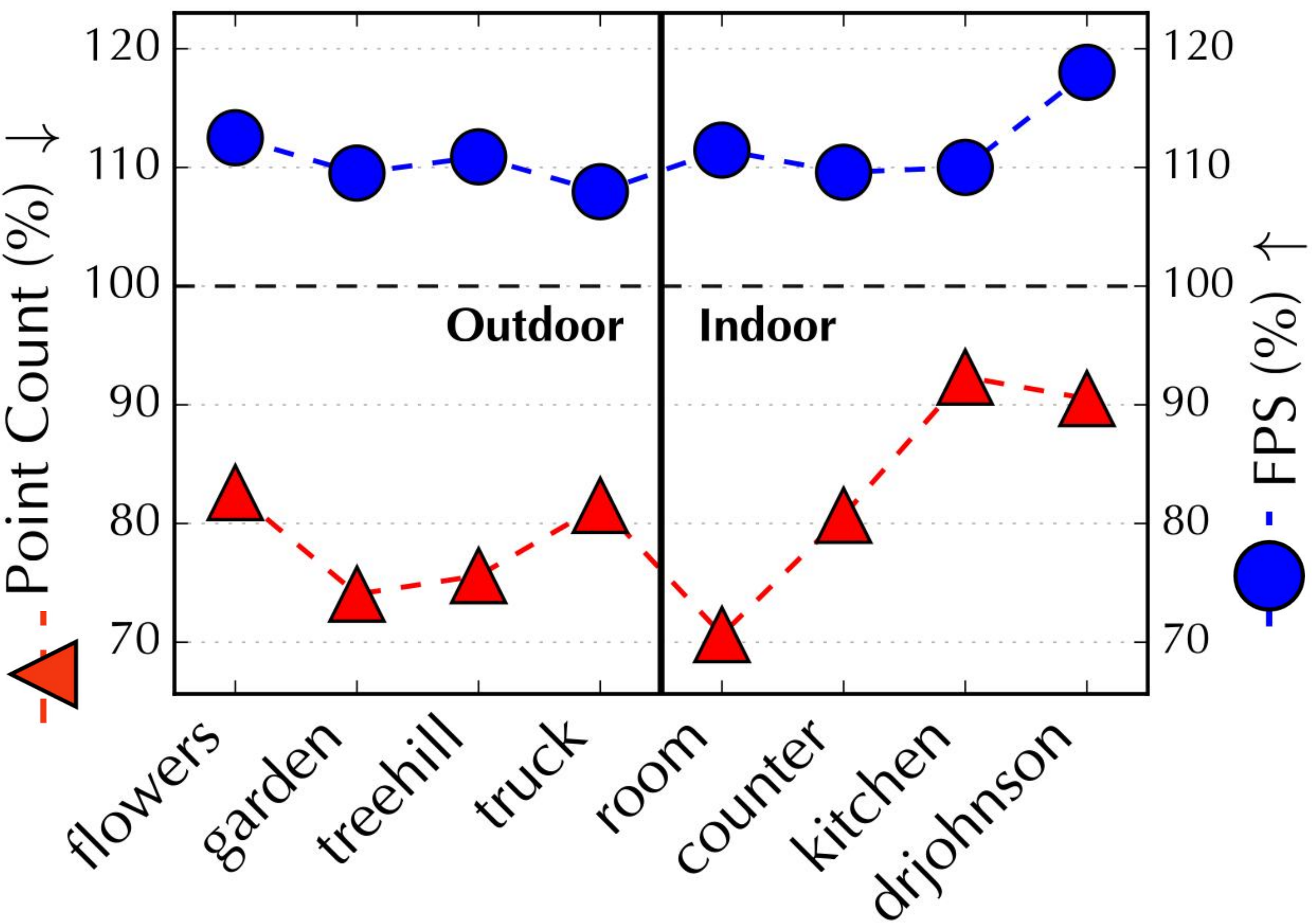}
  }
  \hspace{2em}
  \subfloat[Point Count and FPS Improvement in R4]{
    \includegraphics[width=0.33\textwidth]{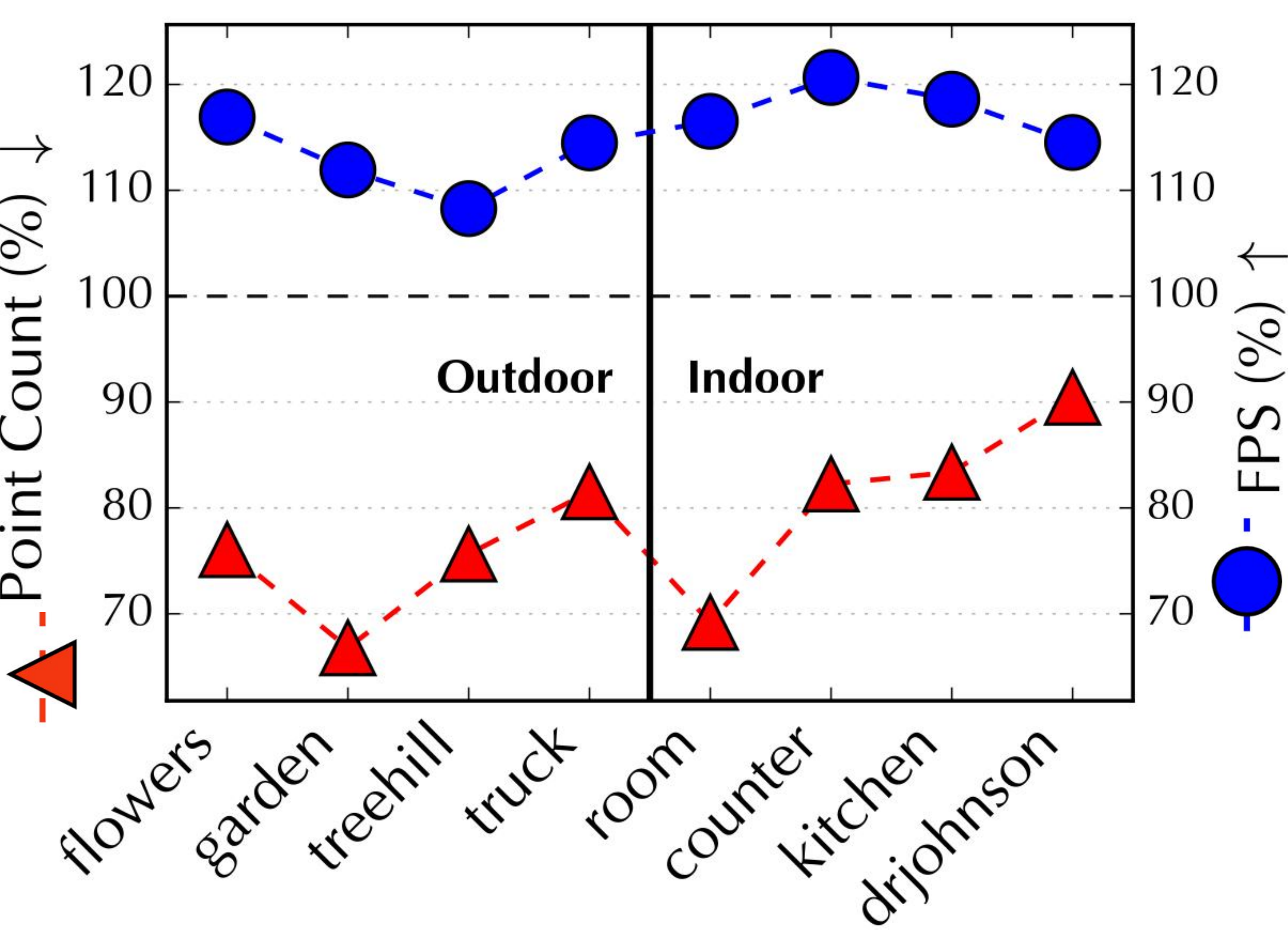}
  }

\caption{
Effectiveness of opacity enhancement (\Sect{sec:fr:enhanced_prim}) in high-eccentricity regions (R3 and R4) across eight scenes. We compare \mode{\newproj-Mono-H} against \mode{\proj-H}.
}
  \label{fig:gaussian_amp_effect}
\end{figure*}

\begin{figure*}[t]
\centering
\begin{minipage}[t]{0.90\textwidth}
  \centering
  \subfloat[Speed-ups of different accelerator variants over the GPU baseline across 9 scenes.]{%
    \includegraphics[width=\linewidth]{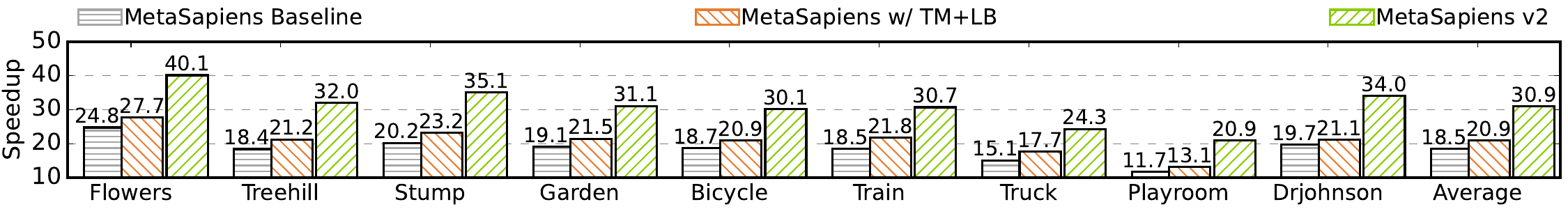}%
    \label{fig:perf_speedup}
  }\\[0.5em]
  \subfloat[Normalized Energy Efficiency of different accelerator variants over the GPU baseline across 9 scenes.]{%
    \includegraphics[width=\linewidth]{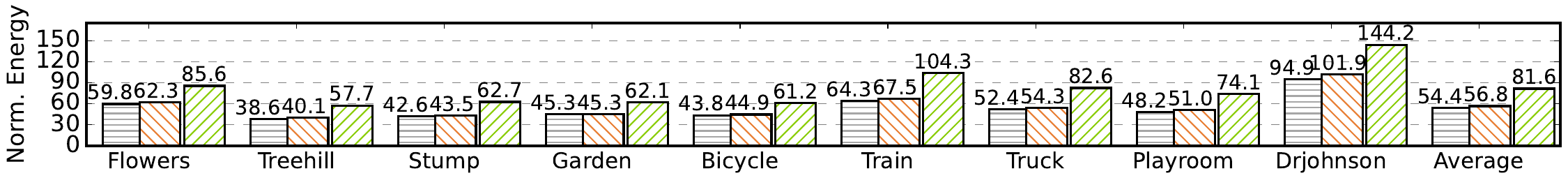}%
    \label{fig:perf_energy}
  }
  \caption{Performance and energy comparison of different accelerator variants.}
  \label{fig:perf}
\end{minipage}
\vspace{-10pt}
\end{figure*}

Our method {offers better speed--quality trade-offs in all metrics.} \mode{\newproj-H-Mono.}, our slowest monocular variant, is 2.0$\times$ faster than the fastest non-FR baseline with better or similar quality. \mode{\newproj-L-Bino.}, our fastest variant, is 8.5$\times$ faster than \mode{3DGS} on average and up to 21.8$\times$ on the largest \texttt{bicycle} trace.

Compared with the strong FR baseline \mode{\proj-H}, \mode{\newproj-H} improves FPS by 7\% and 17\% in monocular and binocular modes while achieving better quality. The modest gain is expected because our techniques mainly accelerate high-eccentricity regions, which are lighter than the fovea.
Opacity enhancement (\Sect{sec:fr:enhanced_prim}) is most effective in sparse, high-eccentricity regions. In \Fig{fig:gaussian_amp_effect}, focusing on R3 and R4, it reduces point count by up to 33\% and improves FPS by up to 20\%. Selective warping is also applied only beyond $18^\circ$ eccentricity (\Sect{sec:warp:method}) to preserve quality, so it primarily accelerates peripheral rendering. Its $\sim$5\% depth-extraction and $\sim$8\% warping overheads, relative to monocular rendering time, also limit the overall binocular speedup to about 10\%.

\subsection{Ablation Study For Algorithmic Designs}
\label{sec:eval:ablation}
{We now ablate the contribution of various performance-enhancing techniques.}
\Fig{fig:fr_comp} shows the FPS (left $y$-axis) and PSNR (right $y$-axis) of: 1) the dense \mode{Mini-Splatting-D} model, 2) \proj with only scale decay (SD; \Sect{sec:prune:scale}), 3) \proj with SD and CE-based pruning (\Sect{sec:prune:metric}), and 4) \proj with SD, CE, and FR (\Sect{sec:fr}).
We use the \mode{\proj-H} model and obtain the FPS/PSNR results on Xavier averaged over all traces. 

The PSNRs for all the variants are similar.
With a similar quality, our SD implementation achieves 1.6$\times$ speedup compared to original dense model;
CE-based pruning and FR bring the speedup to 5.8$\times$ and 7.4$\times$, respectively.
CE reduces the model size by 85\%, and FR diminishes the pruning rate only marginally to 84\% owing to selective multi-versioning.
By incorporating enhanced opacity (\Sect{sec:fr:enhanced_prim}) and selective binocular warping (\Sect{sec:bino}), \newproj further improves \proj performance by 7\% and 17\%, respectively, as discussed in \Sect{sec:eval:render}.

\subsection{Results with Hardware Support}
\label{sec:eval:hw}

\paragraph{Speedup.}
Our hardware support provides further performance improvements.
\Fig{fig:perf} shows the speedups over GPU of: 1) the base accelerator (the uncolored in \Fig{fig:arch}), 2) the accelerator with both Tile Merging and Incremental Pipelining, and 3) the accelerator with binocular rendering support.
We show detailed results from 9 dataset traces.
We use the \mode{\proj-H} and \mode{\newproj-H} for evaluation.
Overall, even the base accelerator achieves a 18.5$\times$ speedup (geomean), up to 24.8$\times$, compared to the GPU baseline across different datasets.

The introduction of Tile Merging and Incremental Pipelining consistently improves performance, because tile merging mitigates the load imbalance across tiles and Incremental Pipelining completely addresses the load imbalance in FR.
Overall, \mode{\proj-TM-IP} combines both techniques and achieves an average 20.9$\times$ (up to 27.7$\times$) speedup.

By introducing selective warping, our method further reduces the rendering workload, as some high-eccentricity tiles can be entirely skipped. This consistently improves speedup. Overall, \mode{\newproj} achieves an average $30.9\times$ speedup, with a maximum of $40.1\times$.

\paragraph{Energy Savings.}
We also show the energy results in \Fig{fig:perf}.
Our base accelerator achieves a 54.4$\times$ energy reduction compared to the GPU baseline. 
\mode{\proj-TM-IP} improves the energy saving to 56.8$\times$;
this is primarily because with incremental pipelining we can afford to smaller SRAMs as line buffers, which reduces the energy consumption of SRAMs.
Finally, \mode{\newproj} boosts savings to $81.6\times$ by replacing a substantial portion of the heavy rasterization workload with lightweight depth blending and warping, reducing both memory access and compute energy.

\subsection{Sensitivity Study for Selective Warping}
\label{eval:ecc_thresh}
\begin{figure}[t]
    \centering
    \includegraphics[width=0.9\linewidth]{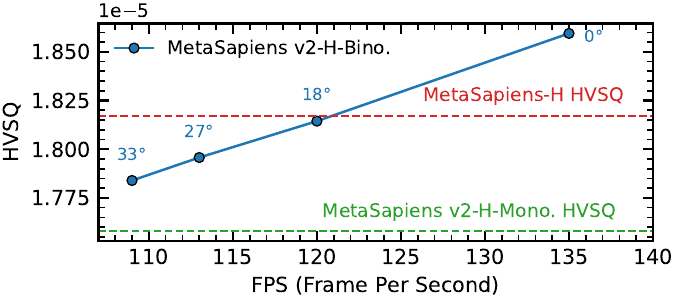}
    \caption{HVSQ--FPS trade-offs across different $\text{Ecc}_{\text{thresh}}$ in our binocular warping. We select $\text{Ecc}_{\text{thresh}} = 18^\circ$ to match the quality of \mode{\proj-H}, which is shown to be no worse than dense PBNR.}
    \label{fig:hvsq_tradeoffs}
    \vspace{-10pt}
\end{figure}

We now examine how $\text{Ecc}_{\text{thresh}}$ affects quality and framerate, justifying its selection.
Since warping is applied only to pixels with eccentricity higher than $\text{Ecc}_\text{thresh}$, a lower $\text{Ecc}_\text{thresh}$ yields higher speedup but larger quality degradation.
\Fig{fig:hvsq_tradeoffs} shows the trade-off between HVSQ (\Sect{sec:bck:hvs}) and framerate for $\text{Ecc}_{\text{thresh}} \in {0^\circ, 18^\circ, 27^\circ, 33^\circ}$ (right to left).
Binocular HVSQ is computed as the average of the directly rendered and selectively warped views.
We also highlight the HVSQ of \mode{\newproj-Mono-H} and \mode{\proj-H}.
Results are averaged over 13 traces across all datasets.

As expected, increasing $\text{Ecc}_{\text{thresh}}$ reduces FPS but improves (lowers) HVSQ.
For any $\text{Ecc}_{\text{thresh}}$, our binocular variants always have higher (worse) HVSQ than monocular ones due to warping.
Among monocular versions, \mode{\newproj-Mono-H} outperforms \mode{\proj-H} due to opacity enhancement, which mitigates background leakage at high-eccentricity and  boosts quality.
We select $\text{Ecc}_\text{thresh} = 18^\circ$ for selective warping, as it aligns the HVSQ of \newproj with that of \mode{\proj-H}, which has subjective quality similar to dense PBNR~\cite{lin2025metasapiens}.

\subsection{Comparison with Other FR Methods}
\label{sec:eval:fr}

Our selective multi-versioning FR (\Sect{sec:fr:rep}) also outperforms the two FR baselines. \mode{SMFR}, a strict-subsetting variant without multi-versioning, is fastest but has poor quality because it sub-samples pre-trained points without fine-tuning; its $L_4$ HVSQ is over 10$\times$ worse than the others. Our method multi-versions only four of about 60 trainable parameters, adding only 6\% storage while reducing the dense model size to 16
\mode{MMFR}, which multi-versions all parameters, improves HVSQ in peripheral regions but is slow and storage-heavy, falling well below the 90 FPS target. Since our FR is already subjectively no worse than dense PBNR~\cite{lin2025metasapiens}, \mode{MMFR} over-optimizes imperceptible details.

\begin{table} 
\caption{Comparison of FR methods.}
\resizebox{\columnwidth}{!}{
\renewcommand*{\arraystretch}{1}
\renewcommand*{\tabcolsep}{4pt}
\begin{tabular}{ c|cccccc } 
\toprule[0.15em]
 \textbf{Methods} &  FPS $\uparrow$  & Storage (MB) $\downarrow$ &  \multicolumn{4}{c}{ HVS Quality ($\times 10^{-5}$)$\downarrow$ }  \\
 &  & & L1 & L2 & L3 & L4 \\
\midrule[0.05em]
SMFR & 125.9 (1$\times$) & 161.6 (1$\times$) & 2.12& 10.1& 21.7 & 28.3\\
MMFR  & 52.6 (0.42$\times$) & 311.0 (1.92 $\times$)  & 2.12 & 1.87 & 1.79 & 1.76\\
Ours & 102.2 (0.81$\times$) & 171.8 (1.06$\times$)  & 2.12 & 2.10 & 2.09 & 2.08 \\
\bottomrule[0.15em]
\end{tabular}
}
\label{tab:fr_comp}
\end{table}

\subsection{Ablation of Hardware Design}
\label{sec:eval:gscore}
\begin{figure}[t]
    \centering
    \includegraphics[width=0.5\linewidth]{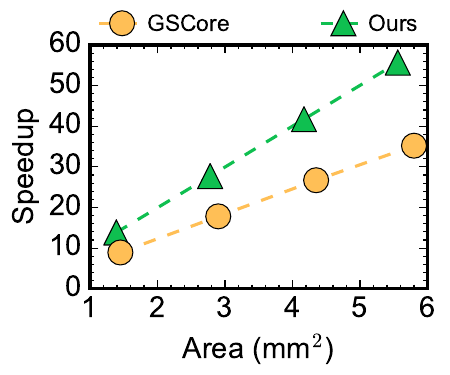}
    \caption{Speedup and area comparison between our hardware and GSCore under different configurations.}
    \label{fig:compare_gscore}
    \vspace{-10pt}
\end{figure}

We now isolate the effect of our incremental pipeline and merging unit (\Sect{sec:hw:load}) by comparing with GSCore.
Our accelerator is based on GSCore~\cite{lee2024gscore}, but has a larger area (\Sect{sec:exp}).
This is because our baseline hardware (uncolored in \Fig{fig:arch}) has 4$\times$ more Volume Rendering Cores compared to that of GSCore with 2$\times$ fewer sorting unit to balance the latency of different stages.
\Fig{fig:compare_gscore} compares speedup over GPU and area between our architecture (with TM and IP) and GSCore, both running \mode{\proj-H} on the \texttt{flowers} scene. We scale both designs proportionally to their own resource ratios. Our architecture consistently achieves higher speedup with slightly smaller area; for example, at around 6 mm$^2$, \proj outperforms GSCore by 1.6$\times$. The gain comes from TM and IP, which reduce tile-pipelining stalls, and becomes larger at higher area where workload imbalance leaves more resources idle.

%% file: related.tex
\section{Related Work}
\label{sec:related}

\paragraph{Foveated Rendering.}
The graphics community has long used FR for real-time rendering~\cite{guenter2012foveated, Chakravarthula2021GazeContingent}. 
Recent work applies FR to neural rendering~\cite{deng2022fov, rolff2023vrs}, such as Fov-NeRF~\cite{deng2022fov}. 
Instead of focusing on NeRF, we target the more efficient PBNR setting. 
Moreover, some methods~\cite{deng2022fov} use a multi-model design similar to our \mode{MMFR} baseline, whereas our subsetting with selective multi-versioning avoids multi-model overhead (\Sect{sec:fr:mot}) and outperforms it (\Sect{sec:eval:fr}). 
We further enhance PBNR primitives for efficient scene representation (\Sect{sec:fr:enhanced_prim}) and support efficient binocular rendering via selective warping (\Sect{sec:bino}).

Conventional FR often uses heuristics such as blurring to relax quality, while recent perception models provide more principled guidance~\cite{walton2021beyond}. We integrate such models into training to demonstrate their practical utility.

\paragraph{Efficient PBNR.}
Most PBNR optimization work focuses on pruning, based on the observation that many points can be removed without quality loss. Existing methods either train masks to remove points~\cite{lee2024compact} or prune points with low numerical contribution to pixel colors~\cite{fan2023lightgaussian, fang2024mini}. Non-pruning compression methods, such as vector quantization~\cite{fan2023lightgaussian} and distillation~\cite{lee2024compact}, have also been explored.

Our work differs in three ways. First, we show that tile intersections, not point count, determine rendering performance (\Sect{sec:prune:perf}), and propose an intersection-aware pruning metric (\Sect{sec:prune:metric}). Second, we introduce scale decay as an orthogonal technique that complements pruning (\Sect{sec:prune:scale}) and jointly improves performance (\Sect{sec:prune:train}). Third, we use warping to accelerate binocular AR/VR rendering (\Sect{sec:bino}).

\paragraph{Enhanced PBNR Primitives.}
Prior works adapt 3DGS-based PBNR primitives~\cite{Kerbl2023GaussianSplatting} for better geometry reconstruction~\cite{huang20242d} or higher quality~\cite{lu2024scaffold}. Unlike these methods, we target foveated rendering, where sparse peripheral points cause background leakage. We introduce opacity-enhanced primitives to mitigate this issue, which rarely arises in uniformly high-quality rendering.

\paragraph{Warping-based Result Reuse.}
Warping is widely used in AR/VR to reuse computation across eyes~\cite{fehn2004depth} or frames~\cite{feng2024cicero, feng2024potamoi}, but typically requires accurate dense depth, which is difficult in real scenes. Our work instead extracts depth directly from PBNR via depth blending, avoiding offline mesh reconstruction and online depth passes, and applies warping only in the periphery, where lower human visual acuity better tolerates noisy depth.

\paragraph{Neural Rendering Accelerators.}
Prior neural-rendering accelerators mostly target NeRF~\cite{lee2023neurex, feng2024cicero}, which PBNR aims to outperform. Recent work accelerates PBNR~\cite{lee2024gscore,feng2024potamoi,lin2025metasapiens,feng2025lumina}, but our work differs in three aspects: it supports FR with minimal hardware changes, mitigates FR-exacerbated PBNR load imbalance, and adds depth blending and selective warping for efficient binocular rendering.

%% file: conclusion.tex
\section{Conclusions}
\label{sec:conc}

We achieve over an order of magnitude speedup over existing PBNR models while maintaining the quality.
{\newproj combines perception-guided pruning, foveated PBNR, opacity-enhanced peripheral primitives, selective binocular warping, and hardware support to deliver order-of-magnitude speedups for PBNR on AR/VR-class devices.}
The speedup comes from: 1) a pruning techniques that directly optimizes for the compute-cost of PBNR, 2) FR and binocular rendering specialized for PBNR, and 3) hardware support addressing the load imbalance in FR PBNR.